\documentclass[longauth]{aa}

\usepackage{graphicx}

\usepackage{txfonts}
\usepackage{placeins}
\usepackage{rotating}

\graphicspath{ {./}{Figures/} }

\usepackage[breaklinks=true]{hyperref}

\makeatletter
\renewcommand*\aa@pageof{, page \thepage{} of \pageref*{LastPage}}

\usepackage{amsmath}	
\usepackage{amssymb}	
\usepackage{natbib}
\usepackage{multirow}
\usepackage{gensymb}

\newcommand{\HII}{\textrm{H~{\textsc{ii}}}}
\newcommand{\mum}{$\mu$m\xspace}
\newcommand{\molh}{H$_2$\xspace}

\usepackage{color}
\definecolor{lightgrey}{rgb}{0.84, 0.84, 0.84}

\definecolor{cbpurple}{rgb}{0.47, 0.37, 0.94}

\begin{document}

	\title{H$_2$ emission in the Orion Bar, NGC7023 and the Horsehead Nebula observed with the James Webb Space Telescope}
	\titlerunning{H$_2$ emission in the Orion Bar, NGC7023 and the Horsehead Nebula} 
	
	\author{M. Zannese\inst{\ref{iff},\ref{ias}}, A. Sidhu\inst{\ref{uwo},\ref{westo}}, A. Guerras\inst{\ref{uwo},\ref{westo},\ref{ias}}, M. Pound\inst{\ref{michi}}, M. Wolfire\inst{\ref{mary}}, A. G. G. M. Tielens\inst{\ref{leiden},\ref{mary}}, E. Peeters\inst{\ref{uwo},\ref{westo},\ref{carl}}, E. Habart\inst{\ref{ias}}, A. Abergel\inst{\ref{ias}}, M. Baes\inst{\ref{gent}}, O. Berné\inst{\ref{irap}}, C. Boersma\inst{\ref{nasa}}, E. Bron\inst{\ref{obs}}, J. Cami\inst{\ref{uwo},\ref{westo},\ref{carl}}, R. Chown\inst{\ref{ohio}},  E. Dartois\inst{\ref{ismo}}, P. Dell'Ova\inst{\ref{ias}}, J. R. Goicoechea\inst{\ref{iff}}, K. D. Gordon\inst{\ref{stsci},\ref{gent}}, P. Guillard\inst{\ref{iap}}, O. Kannavou\inst{\ref{ias}}, K. Misselt\inst{\ref{az}}, N. Monnier\inst{\ref{ias}}, A. Noriega-Crespo\inst{\ref{stsci}}, T. Onaka\inst{\ref{tokyo}},  D. Van de Putte\inst{\ref{uwo},\ref{westo}}, L. Verstraete\inst{\ref{ias}}, A. N. Witt\inst{\ref{toledo}}}
	\authorrunning{M. Zannese, A. Sidhu, A. Guerras et al.}
	\institute{
		Instituto de Física Fundamental (CSIC), Calle Serrano 121-123, 28006, Madrid, Spain \label{iff}
		\\\email{m.zannese@iff.csic.es} \and  Institut d'Astrophysique Spatiale, Universit\'e Paris-Saclay, CNRS,  B$\hat{a}$timent 121, 91405 Orsay Cedex, France \label{ias} \and Department of Physics and Astronomy, University of Western Ontario,
		London, Ontario, Canada \label{uwo}\and
		Institute for Earth and Space Exploration, The University of Western Ontario, London ON N6A 3K7, Canada \label{westo}  \and
		Department of Astronomy, University of Michigan, 1085 South University Avenue, Ann Arbor, MI 48109, USA \label{michi} \and
		Astronomy Department, University of Maryland, College Park, MD 20742, USA \label{mary}  \and Leiden Observatory, Leiden University, 2300 RA Leiden, The Netherlands \label{leiden} \and
		Carl Sagan Center, SETI Institute, 339 Bernardo Avenue, Suite 200, Mountain View, CA 94043, USA \label{carl} \and Sterrenkundig Observatorium, Universiteit Gent, Krijgslaan 281 S9, B-9000 Gent, Belgium \label{gent} \and
		Institut de Recherche en Astrophysique et Plan\'etologie, Universit\'e Toulouse III - Paul Sabatier, CNRS, CNES, 9 Av. du colonel Roche, 31028 Toulouse, France \label{irap} \and
		NASA Ames Research Center, MS 245-6, Moffett Field, CA 94035-1000, USA \label{nasa} \and
		LUX, Observatoire de Paris, Universit\'e PSL, Sorbonne Universit\'e, CNRS, 92190 Meudon, France \label{obs}\and
		Department of Astronomy, The Ohio State University, 140 West 18th Avenue, Columbus, OH 43210, USA \label{ohio} \and     Institut des Sciences Moléculaires d’Orsay, UMR8214, CNRS, Université Paris-Saclay, 91405 Orsay, France \label{ismo} \and
		Space Telescope Science Institute, 3700 San Martin Drive, Baltimore, MD, 21218, USA \label{stsci}  
		\and Sorbonne Universit\'{e}, CNRS, Institut d'Astrophysique de Paris, 98\,bis bd Arago, 75014 Paris, France \label{iap} 
		\and Steward Observatory, University of Arizona, Tucson, AZ 85721-0065, USA \label{az}\and Department of Astronomy, Graduate School of Science, The University of Tokyo, Bunkyo-ku, Tokyo 113-0033, Japan \label{tokyo} \and
		Ritter Astrophysical Research Center, University of Toledo, Toledo, OH 43606, USA \label{toledo}  }

	\abstract
	{Photodissociation Regions (PDRs) are the regions where radiative feedback from massive stars on molecular clouds is dominant. To study these regions, molecular hydrogen (\molh), the most abundant molecule in the interstellar medium, is a handy tool. The \textit{James Webb Space Telescope} (JWST), with its high spatial resolution, sensitivity, and wavelength coverage, provides unique access to the detection and spatial morphology of the \molh rotational (levels with high $J$ detected for the first time) and rovibrational lines.}
	{Our goal is to analyze \molh line emission detected with JWST in several PDRs (the Orion Bar, NGC7023, and the Horsehead Nebula) to constrain the physical structure of the atomic to molecular transition, i.e. the dissociation front (DF), and compare the impact of the ultraviolet (UV) field ($G_0= 100 - 2\times10^4$) and the gas density ($n_{\rm H} = 10^4-5 \times 10^5$~cm$^{-3}$) across different regions.}
	{By leveraging spectro-imaging data from the JWST's NIRSpec and MIRI-MRS instruments, we mapped the spatial distribution of \molh at very high spatial resolution (down to 0.1", $\sim 2\times10^{-4}$~pc, $\sim 40$~au). We then performed a detailed study of the \molh excitation to derive physical parameters. }
	{In the three PDRs, the \molh rotational and rovibrational lines largely dominate the JWST spectrum (they represent about one third of the total lines with luminities of $L_{\rm H_2}(\text{Orion Bar}) \sim 2 \times 10^{-2}$~erg~cm$^{-2}$~s$^{-1}$~sr$^{-1}$, $L_{\rm H_2}(\text{NGC7023}) \sim  10^{-2}$~erg~cm$^{-2}$~s$^{-1}$~sr$^{-1}$ and $L_{\rm H_2}(\text{Horsehead}) \sim 6 \times 10^{-4}$~erg~cm$^{-2}$~s$^{-1}$~sr$^{-1}$). The analysis of their spatial morphology reveals similar filamentary structures across all regions, resulting from the clouds’ corrugated surface or an increase in density. Filament widths vary between PDRs due to local density differences, which change the radiation penetration scale (smaller at higher densities). Spatial shifts between H$_2$ lines (0.5”) are observed and linked to differences in excitation mechanisms (thermalized vs FUV-pumped lines) and the temperature gradient. Despite differences in irradiation conditions at the ionization front (IF), analysis of \molh rotational excitation yields high gas temperatures that are similar across the three PDRs ($T_{\rm gas} \sim 500-600$~K). This is expected for a certain regime of excited PDRs with $G_0(\text{IF})/n_{\rm H} > 0.02$ (such as Orion and NGC7023), where the  $G_0$ at the DFs and the density are similar ($G_0(\text{DF}) \sim 3000$, $n_{\rm H} = 1-3 \times 10^5$~cm$^{-3}$). However, it is surprising for less excited PDRs like the Horsehead Nebula.
		OPRs in the rotational and rovibrational levels differ across the three PDRs. In the Orion Bar and NGC7023, OPR$_{\rm rot} \sim 3$, whereas in the Horsehead OPR$_{\rm rot} < 2.5$. In the Orion Bar OPR$_{\rm rovib} \sim$ OPR$_{\rm rot}$, whereas in NGC7023 and the Horsehead nebula OPR$_{\rm rovib} \sim \sqrt{\text{OPR}_{\rm rot}}$. We also used H$_2$ rovibrational lines and the \molh $0-0$ S(3) line to estimate the visual extinction across the field of view (FOV), yielding markedly different results that constrain both the geometry and the grain composition.}
	{The analysis of the \molh lines detected with JWST in three emblematic PDRs provides unique access to the physical conditions at very small spatial scales. This allowed us to put strong constraints on the gas density, using several diagnostics, the gas temperature, and the complex geometry of the DFs of PDRs. However, due to the complex 3D geometry, \molh emission is observed throughout the FOV and is dominated by the thin UV illuminated layer, making \molh an unreliable tracer of gas temperature in deeper and coolers molecular layers of the PDR. Therefore, \molh emission is not sufficient to trace the possible temperature and density gradient across PDRs and must therefore be associated with other tracers.}

	\maketitle
	
	\section{Introduction}

	A better understanding of star formation and the evolution of interstellar matter relies on the study of Photodissociation Regions (PDRs). Indeed, these regions dominate the infrared (IR) spectrum of galaxies by reprocessing the radiation output of young stars. PDRs re-emit the stellar radiation in the IR-millimeter wavelength regions through gas lines, aromatic infrared bands (AIBs), and thermal dust emission. In addition, PDR emission traces the regions where radiative feedback, one of the main mechanisms that limit star formation \citep{inoguchi_factories_2020}, is dominant. By their intense ultraviolet (UV) field, massive stars contribute to the dispersal of the cloud by heating the gas and the adding momentum. Moreover, the intense stellar Far-UV (FUV) radiation incident on PDRs drives the physics and chemistry of gas and dust \citep[for a review, see, for example,][]{hollenbach_photodissociation_1999,wolfire2022}.

	\molh is the most abundant molecule in galaxies and is a very useful tool for studying PDRs. It is formed on the surface of interstellar grains, where the grains act as catalysts \citep{habart04,bron14,wakelam_h_2017}. Several processes can excite \molh. In dense PDRs, the lowest rotational levels of \molh are populated by collisions \citep[e.g.,][]{Le_Bourlot_1999}.
	More highly excited levels are populated through FUV pumping driven by the intense UV field, either directly via fluorescence or via an IR cascade following the pumping \citep[e.g.,][]{Black_1987,sternberg}. In addition, during their formation on the surface of interstellar grains, \molh can be formed already excited \citep[e.g.,][]{Hollenbach_1971,Hunter_1978,Duley_1986,Le_Bourlot_1995}. Energy equipartition between internal excitation, translation, and grain heating is assumed, and a third of the energy released from the reaction is converted into internal energy. However, the exact branching ratio remains unknown, and the distribution, probably uneven, depends on the conditions in the PDR and the nature of the grains. These molecular processes depend on the gas state (temperature, density...) and the FUV field intensity. Thanks to the low critical densities of the lowest rotational levels,
	these can be thermalized, and \molh can act as a direct thermometer of the medium \citep[e.g.,][]{Le_Bourlot_1999}. In addition, the transitions between \molh levels are of great importance for the heating and cooling of the gas in dense PDRs \citep[e.g.,][]{tielens:85,sternberg,Burton90}. Close to the ionization front (IF), collisional de-excitation of levels populated through FUV pumping increases the gas' kinetic energy, acting as a heating process. Deeper into the PDR, where \molh self-shielding becomes important, collisional excitation of rotational levels followed by radiative emission is a cooling process (i.e., kinetic energy is transferred into radiation). Hence, studying \molh emission provides strong constraints on the physical conditions and the energy balance in PDRs.

	The wavelength coverage and high sensitivity of the \textit{James Webb Space Telescope} (JWST) in the near- and mid-IR allows the detection of hundreds of \molh rotational (high $J$ levels up to $J_{\rm up}=19$ detected for the first time) and rovibrational (up to $v_{\rm up}=8$) lines in PDRs \citep[e.g.,][]{Peeters_2024,Misselt_2025}. In addition, with its high spatial resolution, JWST resolves \molh emission at very small scales (down to 0.1"). These new data allows us to probe variations in the physical conditions in the critical H/\molh transition zones (or dissociation front (DF), where the abundance with respect to H nuclei of \molh equals that of H, $x($H$_2) \approx x(\rm H)$), which remained poorly known but are of fundamental importance for PDR modeling and data interpretation. At the transition zone, the extinction by dust and \molh self-shielding is sufficient for \molh abundance to increase while the gas temperature and UV field intensity are still high enough to excite \molh for it to emit in the IR \citep[e.g.,][]{tielens:85}. Using the rotational lines of \molh, we investigate variations in gas temperature across the DF and examine possible pressure gradients within PDRs. The objective is to determine whether these thin surface layers are sufficiently heated to photoevaporate from the PDR and how this process depends on irradiation conditions and gas density. 
	Finally, these new data provide numerous \molh rotational and rovibrational lines at high spatial resolution, which provide constraints both on the \molh formation processes in warm gas and grains as well as the mechanisms that control the \molh ortho-to-para ratio (OPR). Three emblematic PDRs have been observed with JWST since the starts of its operations: the Orion Bar \citep[as part of the Early Release Science (ERS) Program 1288, PDRs4All,][]{pdrs4all}, NGC7023 and the Horsehead Nebula \citep[as part of the Guaranteed Time Observation (GTO) Program 1192,][]{abergel_jwst_2024,Misselt_2025}. These PDRs have different densities and irradiation conditions (as further described in Sect. \ref{sect:targets}), providing "PDR templates" for future studies. This allows us to study in detail the impact of the environment on \molh excitation and constrain whether \molh is a precise tracer of physical conditions in PDRs.

	In this paper, we present an analysis of \molh emission observed by the JWST in these three PDRs: the Orion Bar, NGC7023 and the Horsehead Nebula \citep[using the results obtained in][]{Zannese_2025}. In Sect. \ref{sect:targets}, we present an overview of the three targets of this study and summarize previous observations. In Sect. \ref{sect:obs}, we present the observations obtained with the Near-Infrared Spectrograph (NIRSpec) and the Mid-Infrared Instrument - Medium Resolution Spectroscopy (MIRI-MRS), and their reduction. In Sect. \ref{sect:spat_morph}, we present an overview of the \molh lines detected in these PDRs and we study their spatial morphology. We also use \molh lines to evaluate the variation of the extinction throughout the Orion Bar and NGC7023, similarly to what was done in \cite{Zannese_2025} for the Horsehead Nebula. In Sect. \ref{sect:excitation}, we analyze the excitation processes of \molh using excitation diagrams at the DF. In Sect. \ref{sect:physcial_conditions}, we use \molh lines to derive physical parameters such as the gas temperature and the gas density across the DFs. Sect. \ref{conclusions} summarizes our main results and conclusions.

	\section{Targets of interest and previous observations}
	\label{sect:targets}
	\begin{sidewaysfigure*}
		\centering
		\includegraphics[width=0.91\textheight]
		{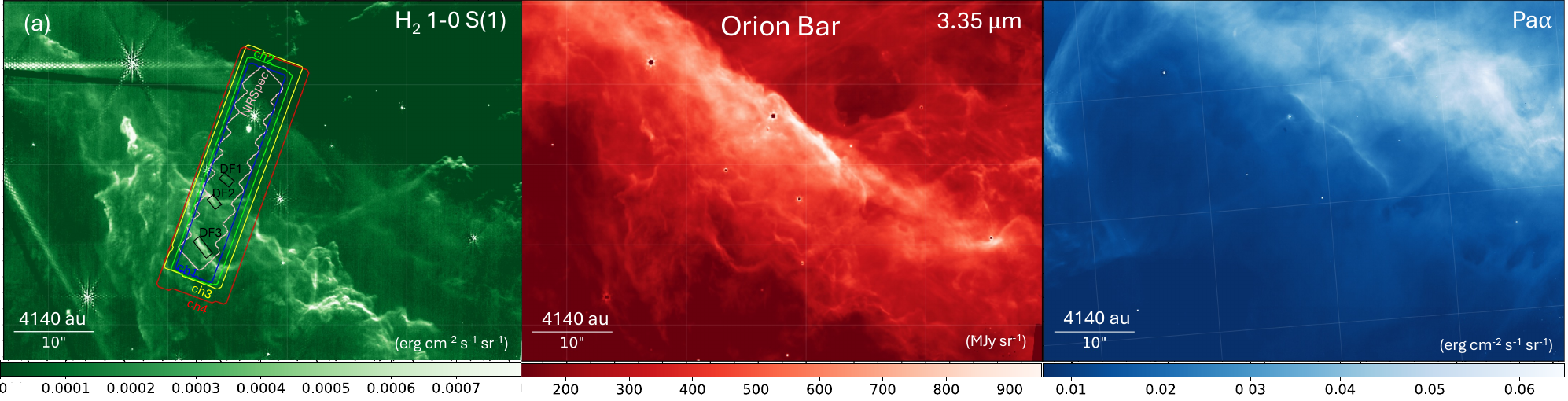}
		\includegraphics[width=0.91\textheight]{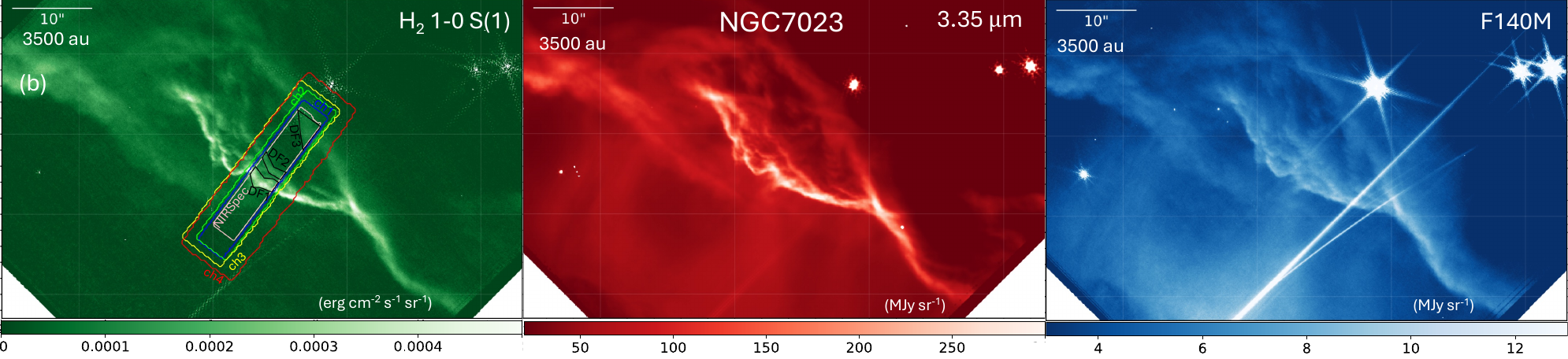}
		\includegraphics[width=0.91\textheight]{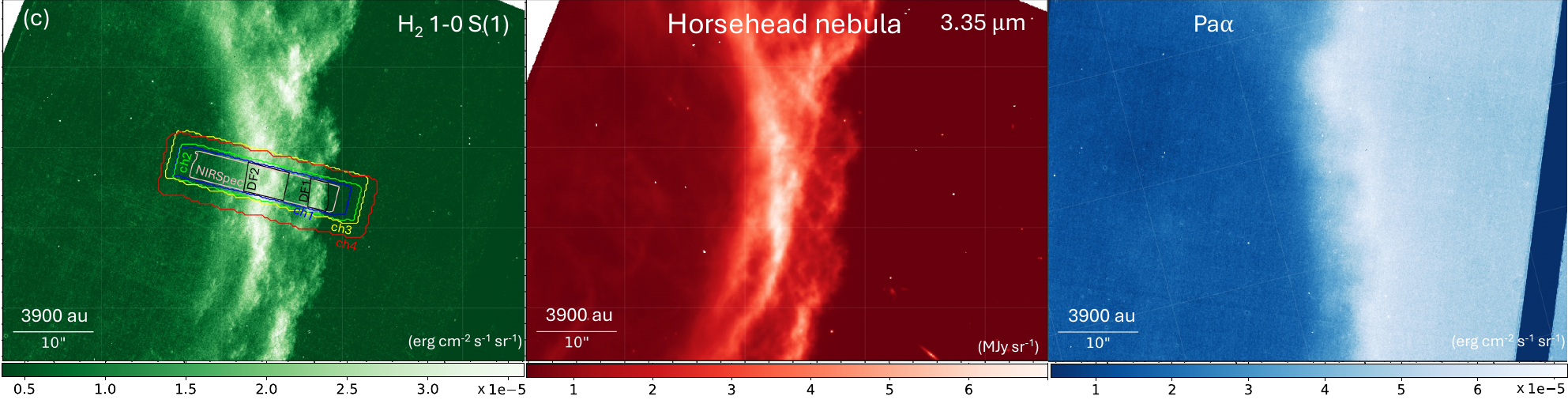}
		\caption{Images of (a) the Orion Bar (data from PDRs4All \citep{Habart_2024} and GTO 1256 \citep{McCaughrean_2023}), (b) NGC7023 \citep[data from GTO \#1192,][]{Misselt_2025}  and (c) the Horsehead Nebula \citep[data from GTO \#1192,][]{abergel_jwst_2024} in various filters. (Left panels) F212N-F210M filter combination (\molh $1-0$ S(1) emission). (Middle panels) F335M filter (Aromatic Infrared Bands). (Right panels) F187N-F182M filter combination (Paschen $\alpha$) for the Orion Bar, F140M (diffuse light) for NGC7023 (no ionized region) and 187N-F210M filter combination (Paschen $\alpha$) the Horsehead Nebula. The field of views (FOVs) of the NIRSpec-IFU (pink) and MIRI-MRS mosaic (channel 1: blue, channel 2: green, channel 3: yellow, channel 4: red). The apertures over the DFs used in this study (black) are shown on top of the H$_2$ map (left panel). The long axes of the IFU FOVs lie along the direction towards the exciting star, towards the upper right (Orion), lower left (NGC7023), and to the right (Horsehead).}
		\label{fig:FOV}
	\end{sidewaysfigure*}
	
	\begin{figure*}
		\centering
		\includegraphics[width=\linewidth]
		{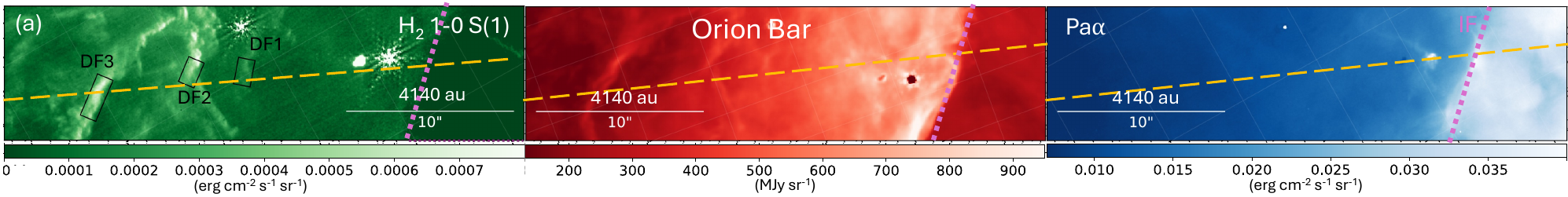}
		\includegraphics[width=\linewidth]{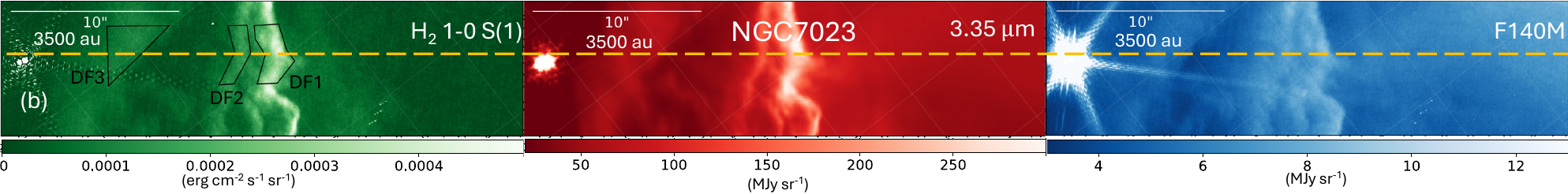}
		\includegraphics[width=\linewidth]{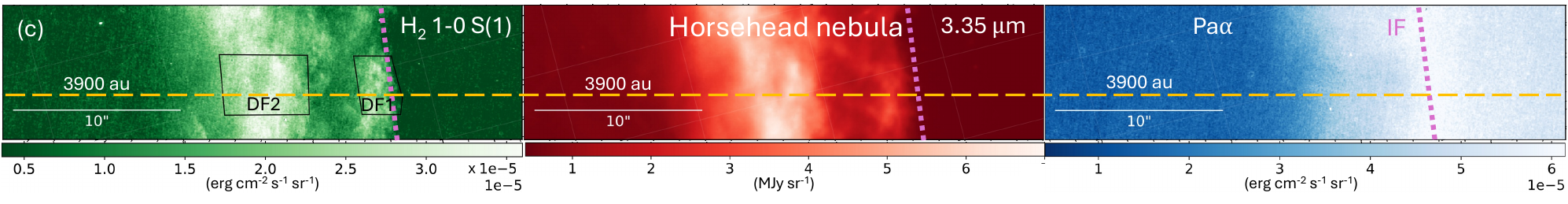}
		\caption{Same as Fig.~\ref{fig:FOV} but zoomed in on the spectroscopic FOV. The yellow dashed lines mark the positions of the cuts used in this study. The purple dotted lines mark the position of the IF. The illuminating star is located on the right.}
		\label{fig:FOVbis}
	\end{figure*}
	\subsection{Orion Bar}
	The Orion Bar, located in the Orion Nebula, the closest site of ongoing massive star-formation \citep[d = 414 pc,][]{Menten07}, is famous for being a highly excited and nearly edge-on PDR. It is irradiated by the intense FUV field originating from the Trapezium cluster, which is dominated by the O7-type star $\theta^1$ Ori C, the most massive star of the Trapezium cluster with an effective temperature $T_{\rm eff} \simeq 39,000$~K. The intense FUV radiation field incident on the IF of the Bar is estimated to be $G_0$(IF) $= (2-7) \times 10^4$ \citep[with $G_0 = 1$ corresponding to a flux integrated between 91.2 and 240 nm of $1.9 \times 10^{-3}$ erg cm$^{-2}$ s$^{-1}$,][]{habing_interstellar_1968} as derived from FUV-pumped IR-fluorescent lines \citep{Peeters_2024}. The main characteristics of this PDR are summarized in Table \ref{tab:carac_PDRs}. Fig.~\ref{fig:FOV}.(a) and Fig.~\ref{fig:FOVbis}.(a) show the Orion Bar through different filters observed with NIRCam. The Orion Bar, being almost edge-on, allows for the observation of the transitions between the ionized gas (in blue), the atomic gas (in red) and the H/\molh transition, or DF (in green).
	
	Previous observations with ALMA \citep[][Atacama Large Millimeter/submillimeter Array]{Goicoechea_2016} and the Keck telescope \citep{Habart_2023} have revealed filament-like structures in the DFs at very small scales, which are also present in the JWST observations \citep{Habart_2024,Peeters_2024}. Several rotational lines were observed with the  \textit{Infrared Space Observatory} (ISO), at lower spatial resolution ($\sim 20$"), from which a rotational temperature $T_{\rm rot} = 390 \pm 20$~K was derived \citep{habart04}. These lines are also observed with the JWST, and the high spatial resolution allows us to recover a higher rotational temperature at the DF ($T_{\rm rot} \sim 600$~K) as the signal is no longer diluted because of the beam \citep[$\sim 1$",][]{van_de_putte_2024}. A more in-depth analysis of these lines is provided in the present paper.  Vibrational levels, probing FUV-pumping, were detected with IGRINS \citep{Kaplan17,Kaplan21} but not in the same region as covered by the JWST observations.

	\subsection{NGC7023}
	
	NGC7023 is often considered an intermediate PDR between a moderately excited PDR and a highly excited PDR. Located in the Cepheus constellation, at $d = 360$~pc \citep{Gaia_2021}, it is irradiated by the $B3Ve-B5$-type binary star HD~200775, with an effective temperature $T_{\rm eff} \sim 18,600$~K \citep{Alecian_2008}. HD~200775 is nestled in the center of a butterfly-shaped cavity that it created from its winds and outflows (now inactive). The cavity edges created by HD~200775 contain a reservoir of PDRs \citep[e.g.,][]{Kohler_2014,bernard-salas_spatial_NGC7023_2015,Joblin_2018}. Among these are the southwestern PDR (SW, 70" southwest of the star), the eastern PDR (E, 170" east of the star), and the northwestern PDR (NW, 40" northwest of the star). The NW PDR is the strongest irradiated of the three with $G_0$(DF) $= 2600$, as derived from the projected distance between the stars and the PDR and an extinction correction of $A_V = 1.5$, \citep[compared to $G_0$(DF) $= 1500$ and $G_0$(DF) $= 250$ for the SW and E PDRs, respectively,][]{Pilleri12} and is oriented nearly edge-on relative to the observer.  The main characteristics of this PDR are summarized in Table \ref{tab:carac_PDRs}. Fig.~\ref{fig:FOV}.(b) and Fig.~\ref{fig:FOVbis}.(b) show NGC7023 NW through different filters observed with NIRCam. This PDR is irradiated by a B star, i.e. a low-ionizing star. Hence, we do not observe a clear ionized region in the field of view (FOV). The region before the DFs (shown in green) is mainly atomic (in red).

	Similar to the Orion Bar, previous high-spatial-resolution observations with the \textit{Canada-France-Hawaii} telescope have revealed complex filamentary structures in \molh emission \citep{Lemaire_1996}. Rotational lines were observed both with ISO \citep{Fuente_2000} and \textit{Spitzer} \citep{Fleming_2010,Boersma2018} and a rotational temperature of $T_{\rm rot} = 522 \pm 35$~K was derived from the \textit{Spitzer} data. Vibrational levels were detected with IGRINS \citep{Huynh_2017} and this study revealed a large density gradient inside the region by analyzing the variation of the $1-0$ S(1)/$2-1$ S(1) ratio (from $10^3-10^4$~cm$^{-3}$ to $10^5$~cm$^{-3}$). Observations with \textit{Spitzer} closer to the irradiating star do not reveal any ionized gas lines nor a clear IF \citep[e.g.,][]{Berne_2012,Mackie_2015}. The widespread H$_{\alpha}$ emission observed in the visible is due to reflected stellar light from the Herbig Be star  \citep{Witt_2006}.

	\subsection{Horsehead Nebula}
	
	The Horsehead Nebula lies at the other side of the excitation scale and is considered a moderately excited PDR. It is located on the western side of the molecular cloud Orion B at a distance of 388~pc \citep{Schaefer_2016}.  It emerges from the edge of the L1630 molecular complex and appears as a dark cloud silhouetted against the \HII\ region IC434 \citep[e.g.,][]{Boer_1983,Neckel_1985,compiegne_aromatic_2007,Pabst_2017,Bally_2018}. The Horsehead Nebula is illuminated by the O9.5V binary system $\sigma$ Orionis \citep{warren_photometric_1977}, which has an effective temperature $T_{\rm eff} \sim 33,000$~K \citep{Simon-Diaz_2015}. The incident UV field on the PDR is estimated to be $G_0$(IF) $ \sim 100$ from the projected distance between the stars and the PDR. The main characteristics of this PDR are summarized in Table \ref{tab:carac_PDRs}. Fig.~\ref{fig:FOV}.(c) and Fig.~\ref{fig:FOVbis}.(c) show the Horsehead Nebula through different filters observed with NIRCam. The imaging data reveal a very small neutral atomic layer \citep[$<100$~au,][]{abergel_jwst_2024}, so the ionized region (in blue) is very close to the DFs (in green). 
	
	High spatial observations of the $1-0$ S(1) line with the \textit{New Technology Telescope} (NTT) revealed bright and narrow filaments at the illuminated edge of the PDR \citep{habart05}. The pure rotational lines have been detected with \textit{Spitzer} and have revealed that only the levels $v = 0$, $J < 5$ are thermalized \citep{Habart2011}, with a rotational temperature around $T_{\rm rot} \simeq 250-400$~K. Many vibrational levels were detected with IGRINS and exhibited a low OPR, characteristic of FUV-pumping \citep{Kaplan21}. PDR models have indicated a steep density gradient at the edge, with a scale length of $\leq 0.02$~pc \citep[from $n_{\rm H} \sim 10^4$~cm$^{-3}$ to $n_{\rm H} \sim 10^5$~cm$^{-3}$,][]{habart05,Goicoechea_2006}. The analysis of \molh rotational and rovibrational lines observed with JWST is presented in \cite{Zannese_2025}, whose results are compared with our study in this paper. They study in detail \molh excitation and show spatial separation from \molh lines excited by different mechanisms (collisions and FUV-pumping). A very high rotational temperature is found across the whole FOV, indicating that \molh lines only trace the 
	illuminated layer and are not reliable tracers of the gas temperature inside the PDR. In addition, this high rotational temperature cannot be reproduced by template stationary, 1D PDR models (see Appendix \ref{appendix:pdrmodels}), highlighting the importance of dynamical effects in this PDR.

	\begin{table*}[!h]
		\caption{Main characteristics of the Orion Bar, NGC7023 and the Horsehead Nebula and parameters derived from \molh lines.}
		\centering
		\begin{tabular}{c|c|c|c}
			Region    & Orion Bar & NGC7023 & Horsehead  \\
			\hline\hline
			Distance (pc)  & 414 $\pm$ 7 \tablefootmark{a} & 355 $\pm$ 5\tablefootmark{b} & 388 $ \pm$ 2\tablefootmark{c} \\\hline
			Illuminating Star & $\theta^1$ Ori C & HD~200775 & $\sigma$ Ori\\\hline
			Spectral Type & O7V & B3Ve-B5 & O9.5V \\\hline
			$T_{\rm eff}$ (K) & 39,000\tablefootmark{d} & 18,600\tablefootmark{e} & 33,000\tablefootmark{f} \\\hline
			Projected distance between the star and the IF (pc) &  $\sim 0.22$ & - &  $\sim 3.7$ \\\hline
			Projected distance between the star and the DF (pc)  & $\sim 0.25-0.27$  & $\sim 0.075-0.085$ & $\sim 3.7$  \\ \hline
			$G_0$(IF) & $(2-7) \times 10^4$\tablefootmark{g} & - & 100\tablefootmark{h} \\\hline
			$G_0$(DF) & 1300-5000\tablefootmark{i} & 2600\tablefootmark{j} & 100\tablefootmark{h} \\\hline\hline
			\multicolumn{4}{c}{Derived in this study from \molh lines} \\\hline\hline
			Temperature (K) & $580 \pm 16$ & $608 \pm 21$ & $506 \pm 19$ \\ \hline
			Column density (cm$^{-2}$) & $1.3 \times 10^{21}$ & $4.7\times 10^{20}$ & $(0.4-2) \times 10^{20}$ \\ \hline
			Density upper limits (cm$^{-3}$) & $(3-6) \times 10^5$ &  $(1-2) \times 10^5$ &  $(1.5-3) \times 10^4$ \\ \hline
			OPR$_{\rm rot}$ & $\sim 3$ & $\sim 3$ & $\sim 2.3$ \\\hline
			OPR$_{\rm rovib}$ & $\sim 2.3-3$ & $\sim 1.2-2.5$ & $\sim 1.1-1.6$ \\
		\end{tabular} 
		\tablefoot{\tablefoottext{a}{\cite{Menten07},}\tablefoottext{b}{\cite{Gaia_2021},}\tablefoottext{c}{\cite{Schaefer_2016},}\tablefoottext{d}{\cite{Simon-Diaz_2006},}\tablefoottext{e}{\cite{Alecian_2008},}\tablefoottext{f}{\cite{Simon-Diaz_2015},}\tablefoottext{g}{\cite{Peeters_2024},}\tablefoottext{h}{\cite{habart05},}\tablefoottext{i}{derived in this work from the $G_0$ value at the IF and estimating the attenuation due to extinction by dust and gas (Sect. \ref{sect:excitation}),}\tablefoottext{j}{\cite{Pilleri12}.}}
		\label{tab:carac_PDRs}
	\end{table*}

	\section{Observation and data reduction}
	\label{sect:obs}
	
	In this paper, we employ both the available NIRSpec-IFU and MIRI-MRS observations to cover the full spectral range of the telescope. The NIRSpec observations cover the 0.97–5.27~\mum range at a spectral resolution of $\sim$2700 while the MIRI-MRS observations cover the 4.9-27.9~\mum range at a spectral resolution of $\sim$3700 in Channel 1 to $\sim$1700 in Channel 4.

	\subsection{Orion Bar}

	The observations of the Orion Bar are part of the PDRs4All ERS Program \citep[PID:1288; ][]{pdrs4all}\footnote{DOI: 10.17909/pg4c-1737}. The $9\, \times 1$ NIRSpec mosaic is centered at $\alpha_{\rm J2000} = 05^{\rm h}35^{\rm m}20.4749^{\rm s}$, $\delta_{\rm J2000} = -05 \degree 25'10.45"$ with a position angle (PA) of 43.74\textdegree\, (Fig.~\ref{fig:FOV}.(a)). This mosaic covers the key zones of the Orion Bar PDR: the \HII\, region, the IF, the atomic region, the DFs, and the molecular cloud. The $1\, \times 9$ MIRI-MRS mosaic overlaps with the NIRSpec mosaic (see Fig.~\ref{fig:FOV}.(a))\footnote{Available on MAST: \url{https://mast.stsci.edu}; DOI: 10.17909/wqwy-p406}.
	The NIRspec data are reduced with the JWST Science Calibration Pipeline (version 1.10.2.dev26+g8f690fdc) and context jwst\_1084.pmap of the Calibration References Data System (CRDS).  
	The MIRI-MRS data are reduced with the JWST Science Calibration Pipeline (version 1.11.1) and CRDS context jwst\_1097.pmap \citep[for further details on the data reduction process, see][]{Peeters_2024,Chown_2024,van_de_putte_2024}.
	To analyze \molh emission in the Orion Bar, we use apertures tracing the DFs in the Orion Bar. The extraction apertures are shown in Fig.~\ref{fig:FOV}.(a) and Fig.~\ref{fig:FOVbis}.(a) and the spectral extraction method is discussed in \citet{Peeters_2024}. 
	
	\subsection{NGC7023}
	
	The observations of NGC7023 are part of the GTO Program \#1192 \citep{Misselt_2025}. We use the NIRSpec-IFU and MIRI-MRS observations centered at $\alpha_{\rm J2000} = 21^{\rm h}31^{\rm m}31.9665^{\rm s}$, $\delta_{\rm J2000} = 68\degree 10'25.00"$ (Fig.~\ref{fig:FOV}.(b)). The IFU mosaics are composed of six pointings to create rectangles, with the long axis oriented along the line of sight when looking from the irradiating star, perpendicular to the PDR front, to cover the different key regions (atomic region, DFs and molecular cloud). The MIRI and NIRSpec data were processed with pipeline version 1.14.0 and CRDS context 1242 and 1241, respectively \citep[for further details on the data reduction process, see][]{Misselt_2025}. Similar to the Orion Bar, we use apertures to trace the DFs. We use those defined by \citet{Misselt_2025}, which are shown in Fig.~\ref{fig:FOV}.(b) and Fig.~\ref{fig:FOVbis}.(b).
	
	\subsection{Horsehead Nebula}
	
	The observations of the Horsehead Nebula are also part of the GTO \#1192 program. They are centered on  position $\alpha_{\rm J2000} = 05^{\rm h}40^{\rm m}53.65^{\rm s}$, $\delta_{\rm J2000} = -2\degree 28'04.39"$. The observations cover a strip across the Horsehead filaments at the interface between ionized and molecular gas (see Fig.~\ref{fig:FOV}.(b)). In this paper, we use the same data reduction as in \cite{Zannese_2025}. The MIRI data reduction was performed with the JWST Science Calibration Pipeline version 1.17.1, using context 1326 for the CRDS, following standard procedures. The NIRSpec data were processed using the JWST science pipeline version 1.14.0 and the CRDS context jwst\_1242.pmap \citep[for further details on the data reduction process, see][]{Misselt_2025}. Here, we also used the apertures defined by \citet{Misselt_2025}, which are shown in Fig.~\ref{fig:FOV}.(c) and Fig.~\ref{fig:FOVbis}.(c).
	
	\begin{figure*}
		\centering
		\includegraphics[width=0.9\linewidth]{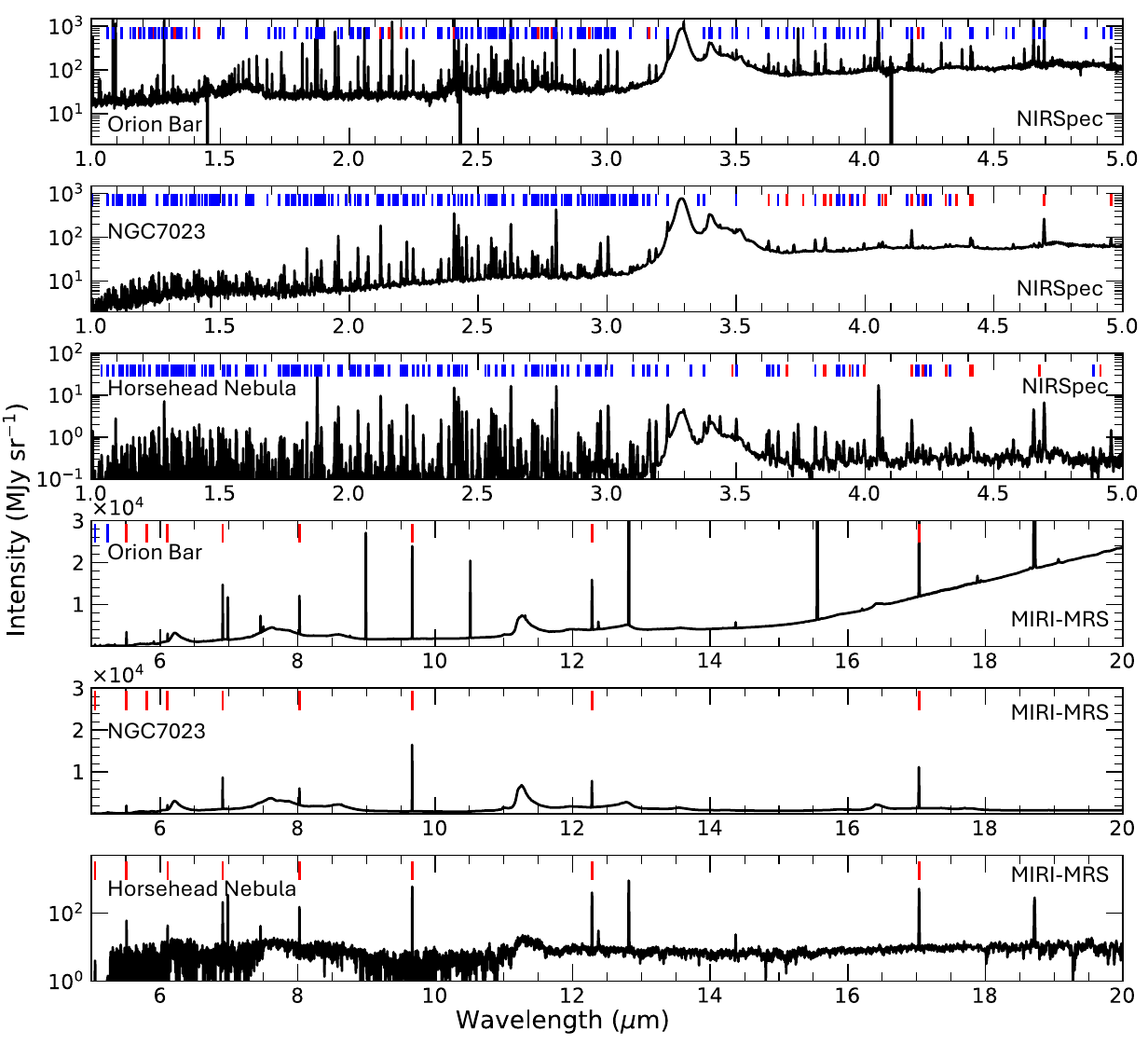}
		\caption{NIRSpec (top) and MIRI-MRS (bottom) spectrum averaged on DF3 in the Orion Bar, DF1 in NGC7023, and DF1 in the Horsehead Nebula. Red lines (resp. blue lines) correspond to the detected rotational transitions (resp. rovibrational transitions) of H$_2$. The identification of other lines can be found in \cite{Peeters_2024} and \cite{van_de_putte_2024} for the Orion Bar and \cite{Misselt_vizier_2025} for NGC7023.}
		\label{fig:full_spec}
	\end{figure*}
	
	\begin{figure*}
		\raggedleft
		\includegraphics[width=\linewidth]{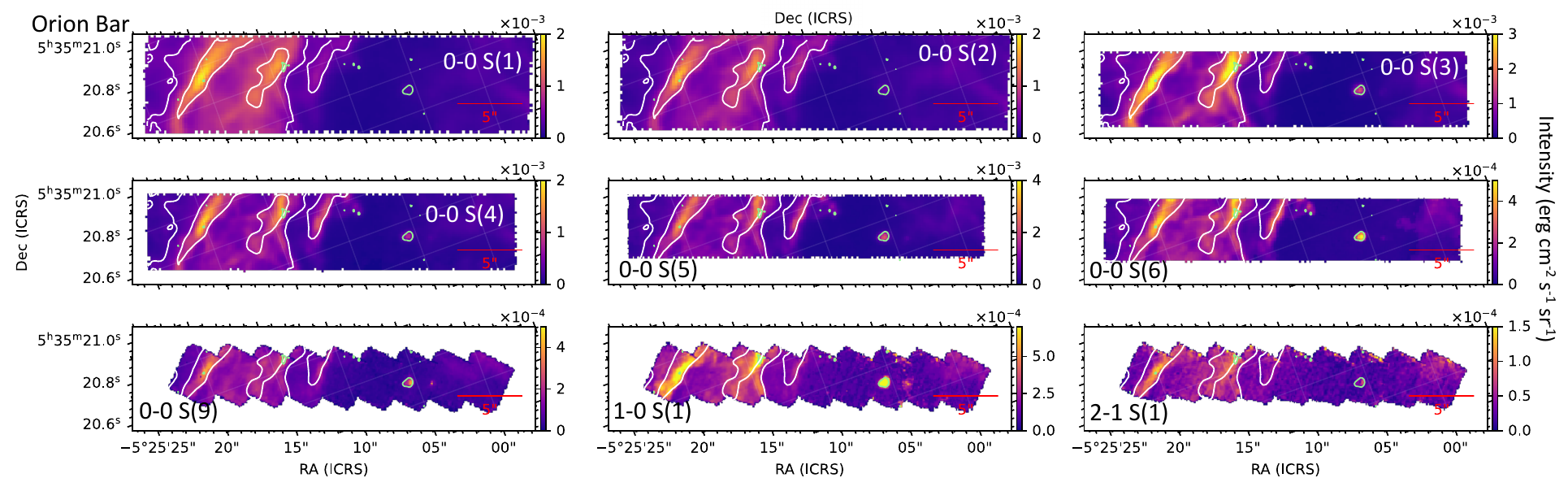}
		\includegraphics[width=0.95\linewidth]{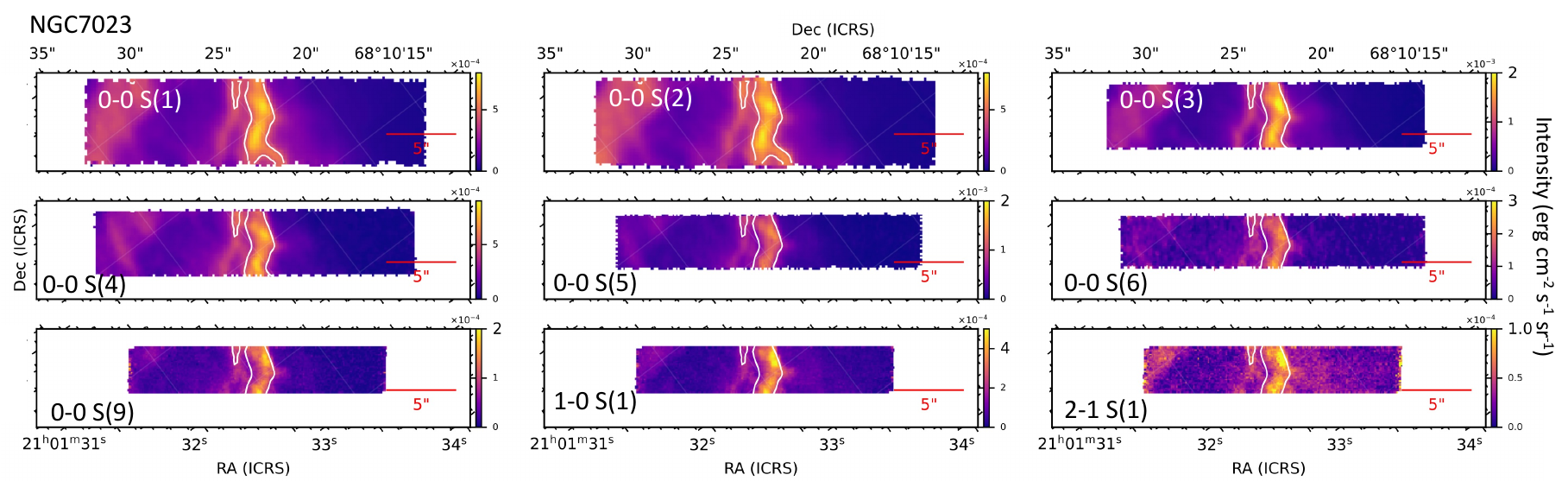}
		\includegraphics[width=\linewidth]{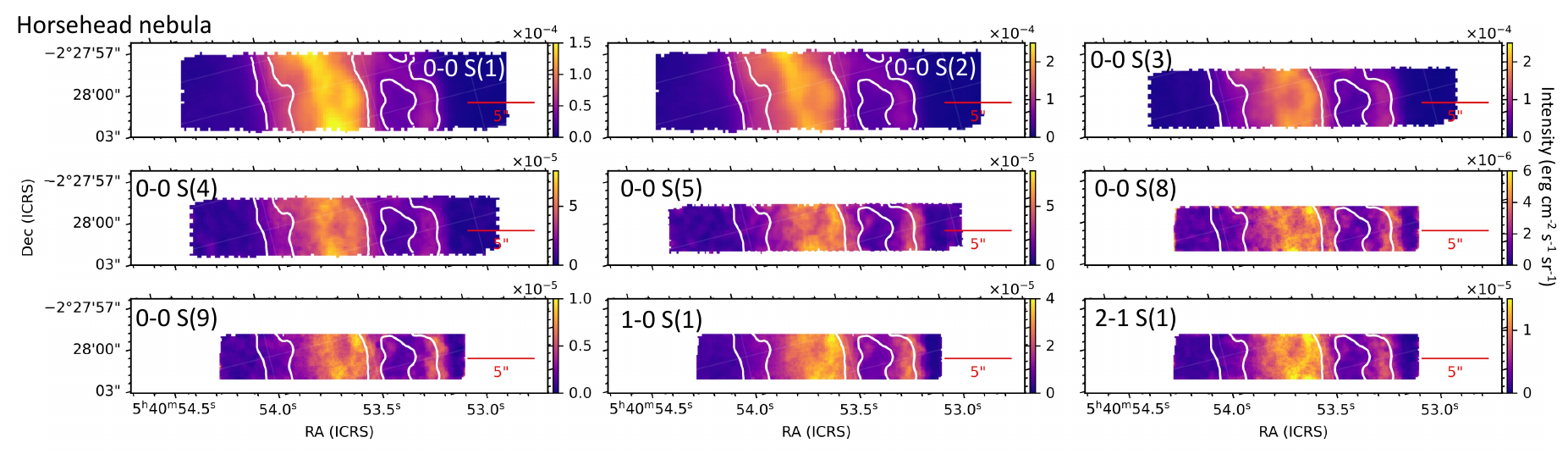}
		\caption{Maps of some H$_2$ rotational line emission and the rovibrational lines 1–0 S(1) and 2–1 S(1) emission obtained with MIRI/MRS and NIRSpec in (top) the Orion Bar, (middle) NGC7023, and (bottom) the Horsehead nebula. White contours are from the $0-0$ S(1) line emission. Green contours focus on the protoplanetary disk d203-506 and are from $1-0$ S(1). The line intensities are not corrected for extinction. In all cases, the illuminating star is located on the right.}
		\label{fig:H2_maps}
	\end{figure*}

	\section{Detection and spatial morphology of \molh emission in the Orion Bar, NGC7023, and the Horsehead Nebula}
	\label{sect:spat_morph}
	\subsection{Detection and measurements}
	
	\label{sec:detection}
	Around 200-300 \molh lines were detected in each PDRs with a total luminosity of $L_{\rm H_2}(\text{Orion Bar}) \sim 2 \times 10^{-2}$~erg~cm$^{-2}$~s$^{-1}$~sr$^{-1}$, $L_{\rm H_2}(\text{NGC7023}) \sim 10^{-2}$~erg~cm$^{-2}$~s$^{-1}$~sr$^{-1}$ and $L_{\rm H_2}(\text{Horsehead}) \sim 6 \times 10^{-4}$~erg~cm$^{-2}$~s$^{-1}$~sr$^{-1}$. We clearly detect lines up to the vibrational levels $v=6$ and potentially a few lines from levels up to $v=8$. Pure rotational lines are observed in $v$ = 0, 1, and 2 levels with lines in rotational levels up to $J=$19, 17, and 9, respectively, in the Orion Bar, $J=$17, 19 and 17 in NGC7023 and $J=$19, 19 and 15 in the Horsehead Nebula. Fig.~\ref{fig:full_spec} displays full NIRSpec and MIRI-MRS spectra averaged in the third dissociation front DF3 of the Orion Bar, the first dissociation front DF1 of NGC7023 and the first dissociation front DF1 of the Horsehead Nebula (see Fig. \ref{fig:FOV}, \ref{fig:H2_maps} and Sect. \ref{sect:spat_morph} for the position in the FOV of these DFs).

	The identification and measurement of the \molh lines were made by \cite{Peeters_2024,van_de_putte_2024} for the Orion Bar, \cite{Misselt_2025} for NGC7023, and \cite{Misselt_2025,Zannese_2025} for the Horsehead Nebula. They all follow a similar method to measure the absolute intensities. They fitted the observed lines with a Gaussian function plus a linear function to account for the continuum, and then integrated the Gaussian function over wavelength. Only the lines with at least a 2-$\sigma$ detection, which were nicely reproduced by the fitting procedure, were kept. To verify this criterion, the residuals were verified in the neighborhood of the line after subtraction of the fit were below or similar to the noise in nearby line-free regions. Identifications and line intensities can be found in the respective papers \citep{Peeters_2024,van_de_putte_2024,Misselt_vizier_2025}.

	\begin{figure*}
		\centering
		
		\includegraphics[width=0.85\linewidth]{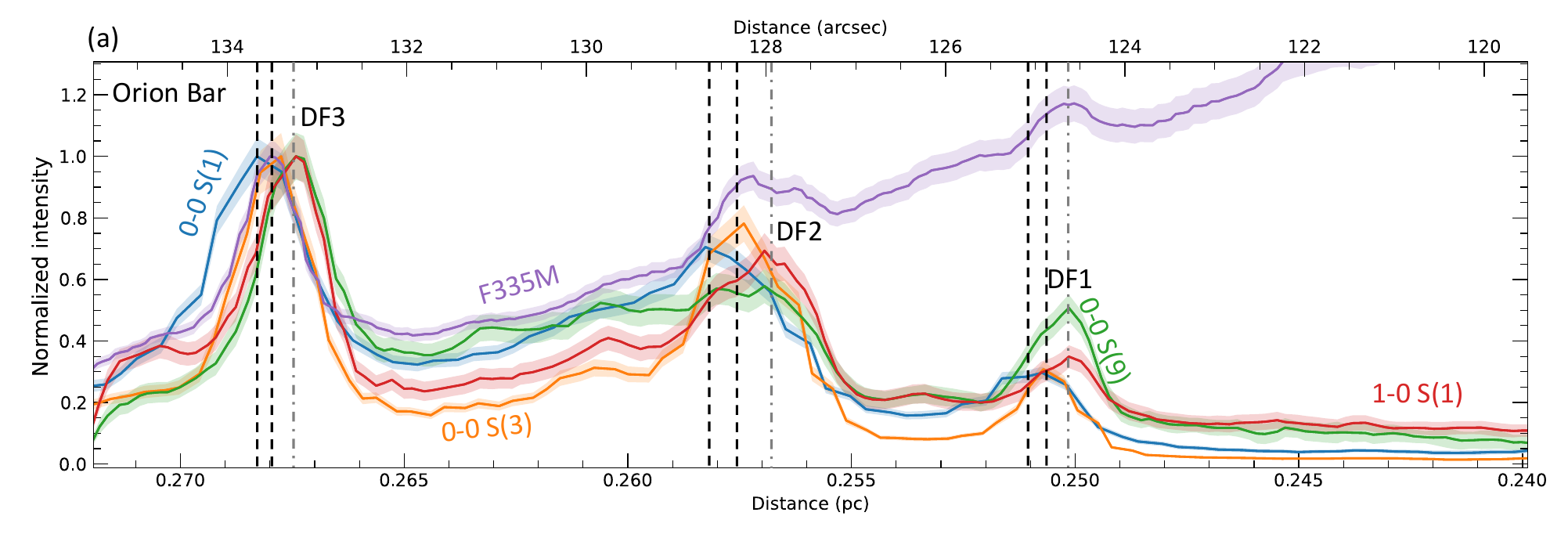}
		\includegraphics[width=0.85\linewidth]{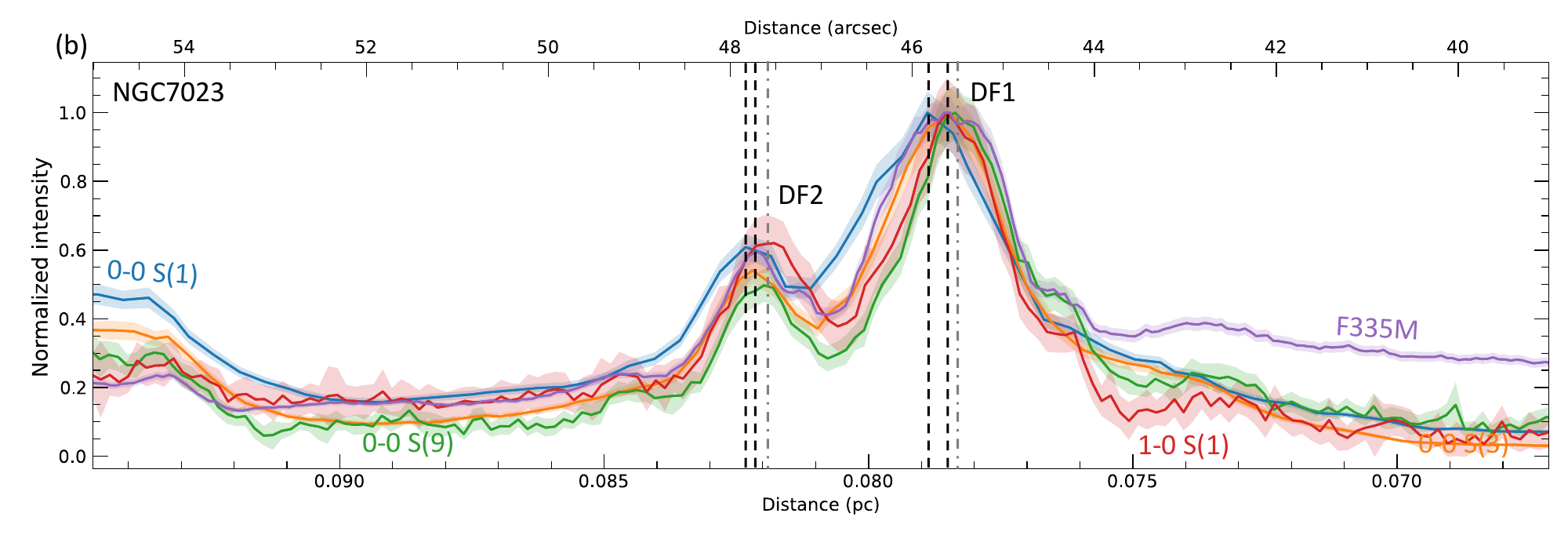}
		\includegraphics[width=0.85\linewidth]{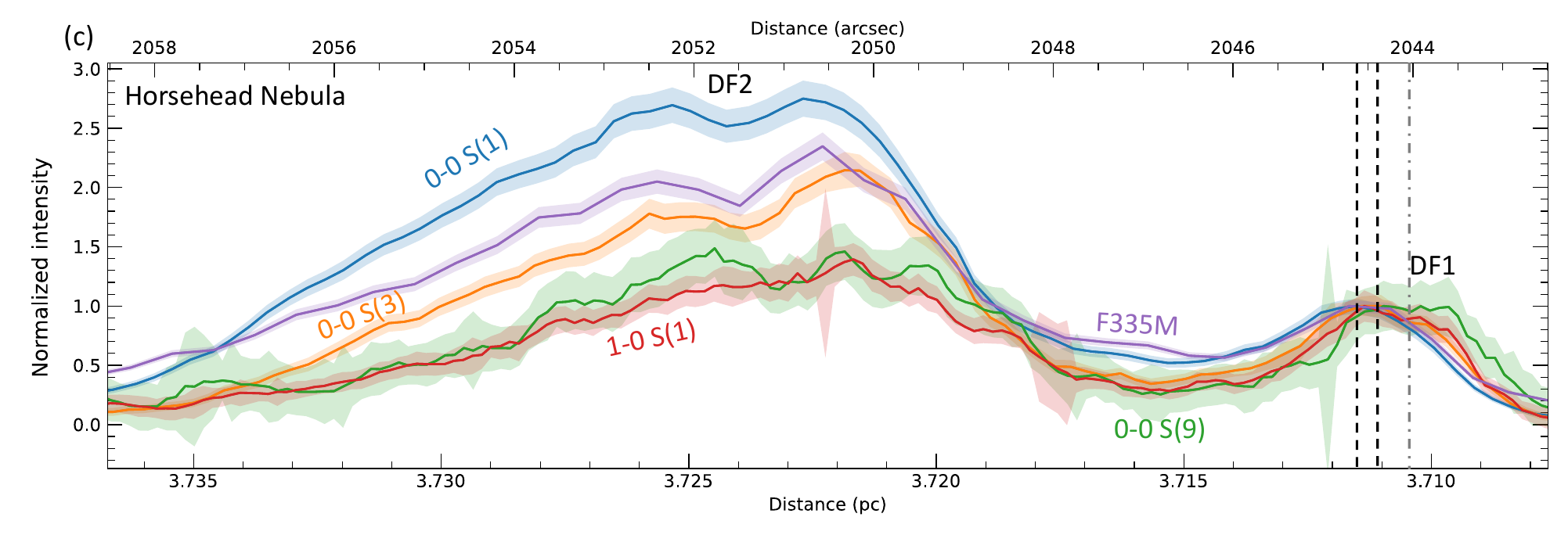}
		\includegraphics[width=0.45\linewidth]{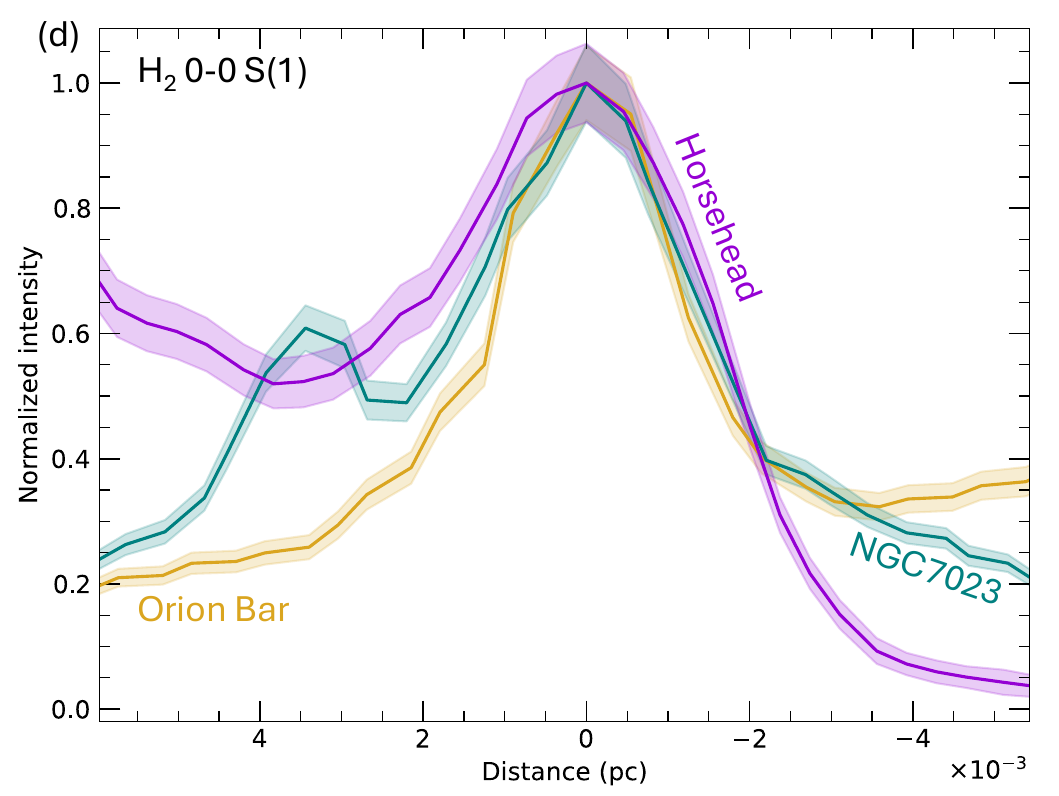}
		\caption{Normalized intensity profiles of some \molh lines and the AIBs from the F335M filter, not corrected for extinction, averaged on 0.5" alongside the inclination of the filaments across the Orion Bar (a), NGC7023 (b), and the Horsehead Nebula (c) as a function of the distance to the illuminating star (pc and arcsec). The dashed black vertical lines mark the position of the thermalized lines' peak, and the dotted-dashed gray vertical lines mark the position of the FUV-pumped lines' peak. (d) Comparison of the \molh $0-0$ S(1) profile centered on the reference DFs (DF3 in the Orion Bar, DF1 in NGC7023 and DF1 in the Horsehead Nebula).    The line intensities are not corrected for extinction.  The illuminating star is located on the right.}
		\label{fig:difference_spatial}
	\end{figure*} 
	\subsection{Spatial morphology of H$_2$ lines}

	Using the IFU capacity of JWST's instruments, MIRI-MRS and NIRSpec, we can study the spatial morphology of \molh emission throughout the PDRs. Using the measurement methods presented in Sect. \ref{sec:detection}, we produce maps of the brightest \molh lines in the three PDRs. Fig.~\ref{fig:H2_maps} displays the maps of a few select \molh lines, 7 rotational lines and two rovibrational lines ($1-0$ S(1) and $2-1$ S(1)) in the Orion Bar, NGC7023 and the Horsehead Nebula.
	The maps were rotated to place the irradiating star on the right-hand side of the figure. In the Orion Bar, we observe three fronts (DF1, DF2, DF3, see Fig.~\ref{fig:FOV}.(a) and Fig.~\ref{fig:FOVbis}.(a)), with DF3 being the brightest, in not extinction-corrected maps, and DF1 the faintest. This complexity can be explained by a terrace-field-like structure \citep[see Fig.~5 of][and Fig. \ref{fig:geometry} in the Appendix]{Habart_2024} or several filamentary structures whose lengths are significantly greater than their widths due to increases in density \citep[see the similar filamentary structure of C$^{18}$O emission,][]{Goicoechea_2025}. To compare the three PDRs simply, template spectra of the DFs are essential. As we expect the DFs to have similar physical conditions and hence similar spectra; in each PDR, we use the one that is easiest to study, considering different criteria: the least extinguished (so that we have less uncertainty due to extinction correction) and/or the brightest (to detect as many lines as possible with good signal-to-noise ratio, SNR). In the rest of the study, for the Orion Bar, we use the brightest, i.e. the third, dissociation front DF3 as a template for all DFs. In NGC7023, we observe two bright filaments in the center of the FOV (DF1 and DF2, see Fig.~\ref{fig:FOV}.(b) and Fig.~\ref{fig:FOVbis}.(b)), with DF1 being the brightest (see the schematic view in Fig. \ref{fig:geometry} in the Appendix). We observe a third filament, at the edge of the maps (on the left, the farther away from the star) that is wider and fainter than DF1 and DF2, so, the physical conditions may differ from those of the other two (see the schematic view in Fig. \ref{fig:geometry} in Appendix). As it is on the edge of the FOV of the observations, we will discard it for the remainder of this study and use the first dissociation front, DF1, as a template for NGC7023. These structures are very similar to what is observed in the Horsehead Nebula, where three dissociation fronts were detected \citep{Zannese_2025}; one located at the very edge of the PDR (DF1) and the two others lying on top of a single filament separated by ~2". For the Horsehead, we follow \cite{Zannese_2025} and use the DF1 region as the template, as it is less affected by extinction. While the structures are similar between the three PDRs, the \molh emitting structures appear narrower in both the Orion Bar and NGC7023, spanning <~1" ($2 \times 10^{-3}$~pc) compared to ~1.5" ($2.9 \times 10^-3$~pc) in the Horsehead (see Fig. \ref{fig:difference_spatial}.(d)). This observation provides the first evidence that the Orion Bar and NGC7023 may be more dense than the Horsehead Nebula. Indeed, in a denser environment, the UV field is attenuated over a shorter distance, and thus the \molh emission is less extended. For NGC7023, a method of data fusion between the MIRI Imager and MIRI-MRS allows us to recover \molh maps at higher spatial resolution, especially at longer wavelengths, such as for the  $0-0$  S(1) line (recovering the smallest Point Spread Function (PSF) of MIRIM, $\sim 0.3$", even at long wavelengths). Appendix \ref{appendix:fusion} presents briefly the method, and Fig. \ref{fig:H2mapsfusion} displays the maps of some of the \molh lines. The detailed method of data fusion is presented in Monnier et al. in prep. 
	
	In these PDRs, we also observe a spatial shift between the different \molh lines. Highly excited lines ($v=0$, $J_{\rm up} > 7$ and $v>0$) peak closer to the edge of the PDR than the lower excitation ones. Fig.~\ref{fig:difference_spatial} shows the normalized intensity profiles of several \molh lines across the Orion Bar, NGC7023, and the Horsehead Nebula along the cuts presented in Fig.~\ref{fig:FOVbis} and averaged over 0.5" along the inclination of the filaments.
	This figure shows that the FUV-pumped lines ($v=0$, $J_{\rm up} > 7$ and $v>0$) peak at the same position, the closest to the edge of the PDR, where the UV field is stronger, and then thermalized lines ($v=0$, $J_{\rm up} \leq 7$) peak deeper into the PDR, where the density is higher. 
	In addition, shifts between pure vibrational $0-0$ S(1) and $0-0$ S(3) peaks are apparent, indicating a temperature gradient. Indeed, the energy of the upper level of $0-0$ S(3) ($E_{\rm up} = 2503$~K) being higher than that of $0-0$ S(1) ($E_{\rm up} = 1015$~K), the $0-0$ S(3) line peaks at a position where the temperature is higher, hence closer to the edge.
	
	\subsection{Evaluation of the attenuation in the line of sight by foreground matter and matter inside the PDR using \molh lines}
	\label{sect:extinction} 
	As \molh lines fall at IR wavelengths, they can be affected by attenuation along the line of sight due to foreground material and material inside the PDR. Then, it is necessary to correct for attenuation for an accurate analysis. Here, we neglect the contribution of the emitting layer of H$_2$ (due to the inclined geometry) to the attenuation and only consider the contribution of the foreground material. Indeed, assuming an angle to the normal of the PDR of 60°, we estimate that the H$_2$ emitting layer contribution to the attenuation is at most $A_V \sim 0.2-0.5$, but this value is very uncertain as the geometry of the PDRs is not well constrained. 
	
	Following the method of \cite{Peeters_2024,Zannese_2025}, \molh emission can be used as a tracer of the attenuation in the line-of-sight. When \molh emission is optically thin, the line intensity is directly proportional to the column density of the upper level.
	Hence, the ratio of the intensities of two lines coming from the same upper level (such as $1-0$ S(1) and $1-0$ O(5)) depends only on the wavelengths and the Einstein coefficients, and is independent of physical conditions. The difference between the observed and theoretical ratio is then solely due to the attenuation along the line of sight. 
	
	Using Eq. 4 of \cite{Zannese_2025}, we derive the attenuation over the entire FOV using 24 line ratios for the Orion Bar and 16 line ratios for NGC7023. These transitions cover wavelengths from 1.1 - 4.3 $\mu$m. We use the total-to-selective extinction $R_V$-parametrized curve from \cite{gordon2023} \citep{Gordon_2009,Fitzpatrick_2019,Gordon_2021,Decleir:22} evaluated at $R_V = 5.5$, representative of a medium with a larger mean size for the grains \citep[e.g.][]{Bohren_1983}, consistent with the depletion of nanograins, which is more adapted for PDRs with a strong incident UV field \citep[e.g.,][]{Cardelli89,Schirmer2022,elyajouri_2024}.
	
	\begin{figure}
		\centering
		\includegraphics[width=0.95\linewidth]{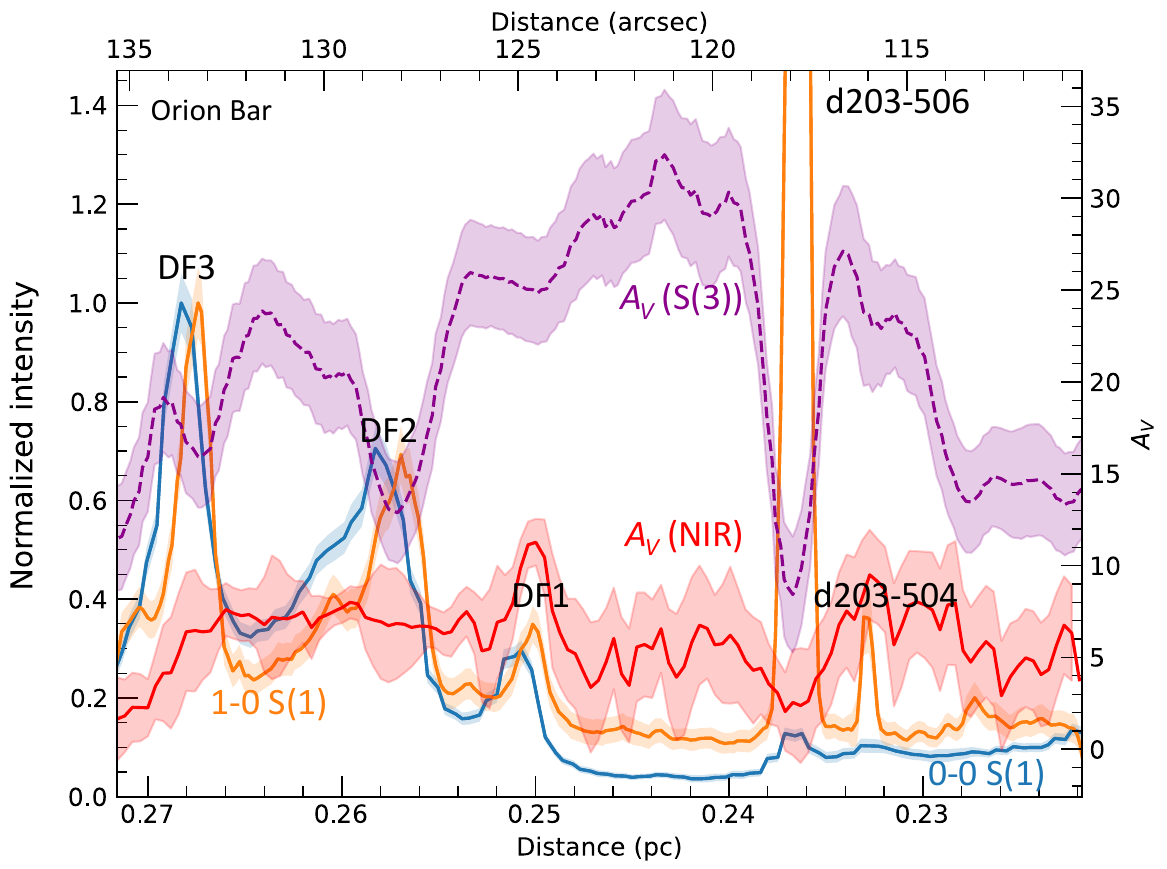}
		\includegraphics[width=0.95\linewidth]{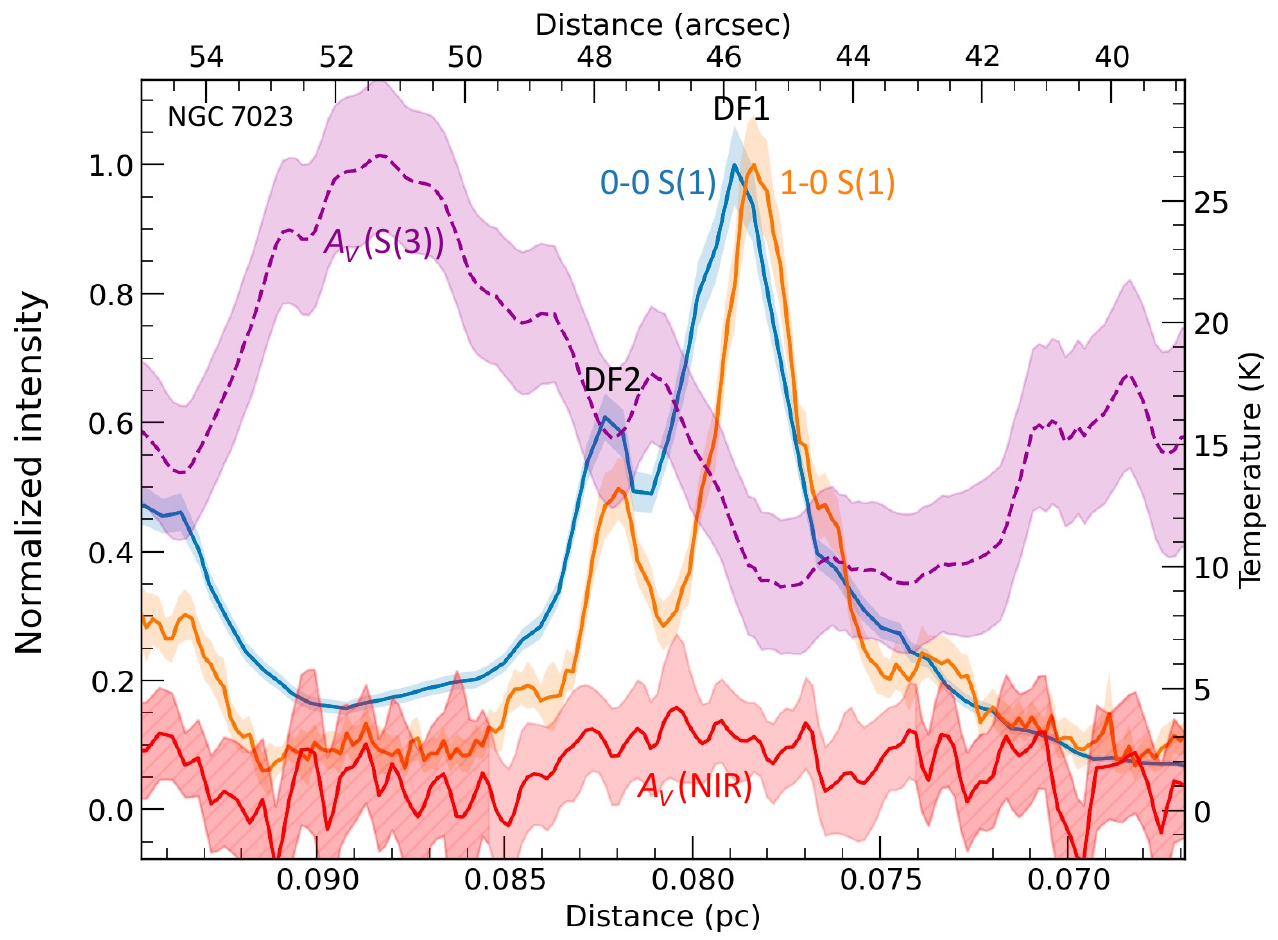}
		\caption{Visual attenuation profiles derived from \molh line ratio in the NIR (red) and using the intensity of the $0-0$ S(3) line (purple) compared to the \molh $0-0$ S(1) and $1-0$ S(1) line emission profile in the Orion Bar (top panel) and NGC7023 (bottom panel)  as a function of the distance to the illuminating star (pc and arcsec). The hatched parts correspond to the regions where \molh rovibrational lines are not detected with enough SNR and, hence, where the estimation is very uncertain. The uncertainties are estimated solely from a statistical perspective (based on the standard deviation of $A_V$ estimates) and do not account for the absolute uncertainties in the intensities. The difference between $A_V$ derived from the S(3) line and NIR lines can be explained by the extinction curve having a shape not adapted to these environments (see Sect. \ref{sect:extinction} for more details).}
		\label{fig:AV_cut}
	\end{figure}

	Fig.~\ref{fig:AV_cut} displays the spatial variation of the visual attenuation in the line of sight across the Orion Bar and NGC7023 \citep[for the Horsehead Nebula, see Fig.~5 of][]{Zannese_2025}. For the Orion Bar, we find again the terrace-field-like structure deduced by \cite{Habart_2024} and \cite{Peeters_2024}, where the extinction toward DF1 ($A_V \sim 10$) is higher than that toward DF3 ($A_V \sim 5$), revealing that DF1 is located further from the surface into the PDR in the line of sight. This likely means that a significant fraction of the \molh emission toward DF1 does not come from the Bar but from the irradiated surface of the background OMC-1 cloud. We observe a plateau between DF1 and DF3, where the attenuation is constant, followed by a decrease after DF3. We also observe that the attenuation is lower towards the disk d203-506, indicating that it is not embedded in the atomic region but is likely between it and the observer, in agreement with \cite{Haworth_2023}. In NGC7023, we observe a relatively uniform, low attenuation across the PDR with $A_V \sim 2-3$. The SNR of excited \molh lines is dropping drastically outside the DFs, making the estimation of the attenuation very uncertain outside of those. However, the attenuation remains very low across the FOV of NGC7023 and should not significantly affect the results of this study, mainly based on MIR lines, which are less affected by attenuation. We present in Fig.~\ref{fig:AVmap} the maps of visual attenuation. 
	
	Interestingly, the evaluation of attenuation cannot account for the weaker intensity of the $0-0$ S(3) line at $\lambda \sim 9.665$~$\mu$m  (see excitation diagram corrected for attenuation in Fig.~\ref{fig:diag_rot}). This means that the extinction curve used in this study underestimates the attenuation around 10 $\mu$m where there is a silicate absorption band. To appreciate how the extinction curve could be affected, we also derived the difference in visual attenuation from the $0-0$ S(3) line following the method described in the Sect. 4.3.4 of \cite{van_de_putte_2024}. To estimate $A_V$(S(3)), we fitted the Boltzmann distribution of the $J=3-7$ levels from the maps not corrected for extinction, excluding $J=5$ ($0-0$ S(3)), and we derived the column density of the $J=5$ level from this fit, assuming an OPR of 3 (see Sect. \ref{sect:excitation}). Comparing the calculated column density to the observed one (not corrected for extinction), we obtain $A_\lambda$(S(3)) $- A_\lambda$(others). Considering the extinction curve of \cite{gordon2023} with $R_V =  5.5$, we have $A_\lambda$(S(3))$/A_V \simeq 0.095$, while $A_\lambda$(others)$/A_V \simeq 0.045$. Hence, we can estimate $A_V$ using $A_\lambda$(S(3)) $- A_\lambda$(others) $\simeq 0.095A_V - 0.045A_V = 0.05A_V$. This method cannot be easily applied to the Horsehead Nebula, as the OPR is not at equilibrium and varies across the PDR \citep{Zannese_2025}. Hence, there would be too many free parameters and the estimation of extinction using the S(3) lines would be very uncertain. 
	For the Orion Bar and NGC7023, we compare the results with that obtained using NIR \molh lines. The spatial variation across the PDRs is shown in Fig.~\ref{fig:AV_cut} and the map is displayed in Fig.~\ref{fig:AVmap_S3}. The overall spatial morphology of the attenuation derived from the NIR lines and $A_V$(S(3)) is very different. Indeed, in the Orion Bar, the attenuation derived from the S(3) line is larger and reaches a maximum in the atomic region. In NGC7023, it increases after the DF, contrary to the attenuation derived from NIR lines. However, in the case of NGC7023, the estimation of $A_V$ outside the DF is very uncertain based on the NIR lines. The difference in both the absolute value of $A_V$ and the spatial morphology of the attenuation in the line of sight reveals that the extinction curve must be different from the one used in this study, but also that there is a change in its shape across the PDR. As discussed in \cite{Peeters_2024} (see their sect. 6.1), this could be due to the illuminating source and the attenuating material being extended and distributed heterogeneously, which has an impact on the amount of scattering in and out of the line of sight and thus the absolute extinction (the summation of absorption and scattering out of the line of sight). Due to the complex geometry, it is challenging to constrain this contribution, but in general, the attenuation will be reduced by the light scattered into the line of sight \citep{Code_1973}. The attenuation in the MIR, always being underestimated using the \cite{gordon2023} extinction curve, shows that the extinction curve must be flatter in the near-IR. In addition, in the case of the Orion Bar, the difference between the two methods being larger in the atomic region than in the DFs reveals that the extinction curve must be even flatter near the IF than deeper inside the PDR. This result is in agreement with the study by \cite{elyajouri_2024}, which find that the nanograins must be strongly depleted in the atomic region with an abundance (relative to the gas) 15 times less than in the diffuse ISM and, thus estimates a very flat extinction curve. Nevertheless, the S(3) line is still not properly corrected when compared to the other rotational lines when using the $A_V$(S(3)) value. This means that either the true value of $A_V$ is higher by about a factor of 1.2 but still has the same spatial variation (as estimated by deriving $A_V$(S(3)) using the same method iteratively) and/or that the strength of the silicate band is underestimated by the \cite{gordon2023} extinction curve. However, understanding the precise shape and variation of the extinction curve is outside the scope of this paper. As the lines that are the most affected by attenuation are in the NIR, we decided to use the attenuation derived with NIR lines to correct all \molh line intensities. As the evaluated attenuation in the MIR remains rather low, most of the \molh rotational lines are only weakly affected by the uncertainty on $A_V$. We choose to discard the $0-0$ S(3) line for the rest of the analysis as it falls in the silicate absorption feature.

	\section{Excitation mechanisms}
	\label{sect:excitation}
	\subsection{Shape of the excitation diagram}
	\label{sect:diag_rot}
	\begin{figure}
		\centering
		\includegraphics[width=1\linewidth]{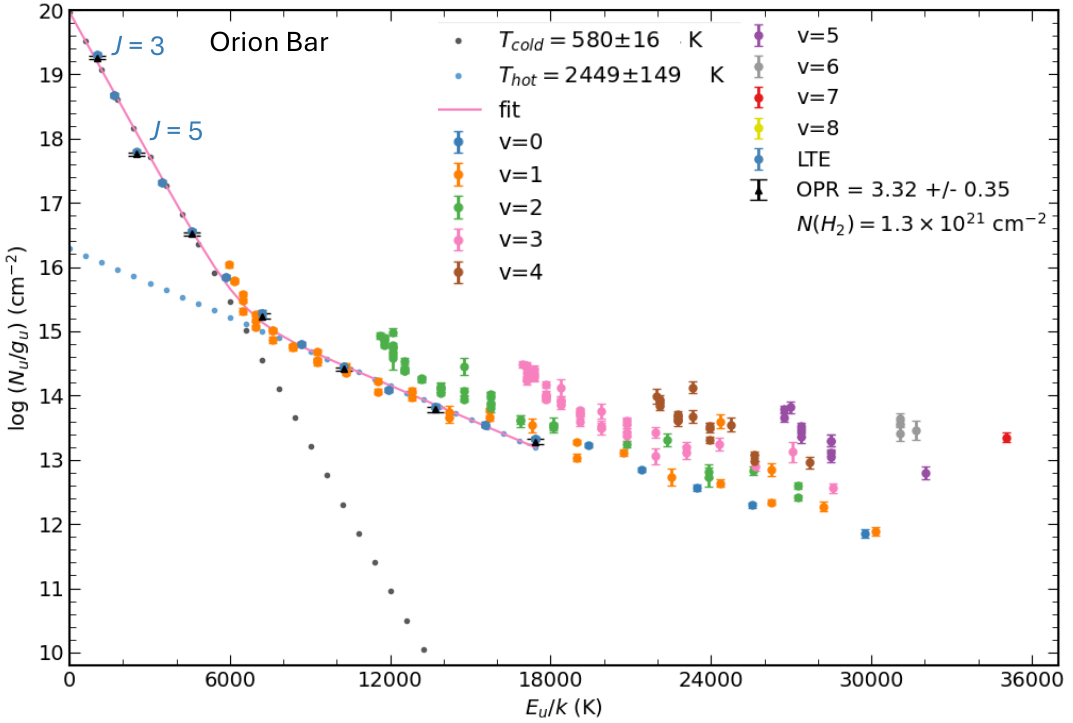}
		\includegraphics[width=\linewidth]{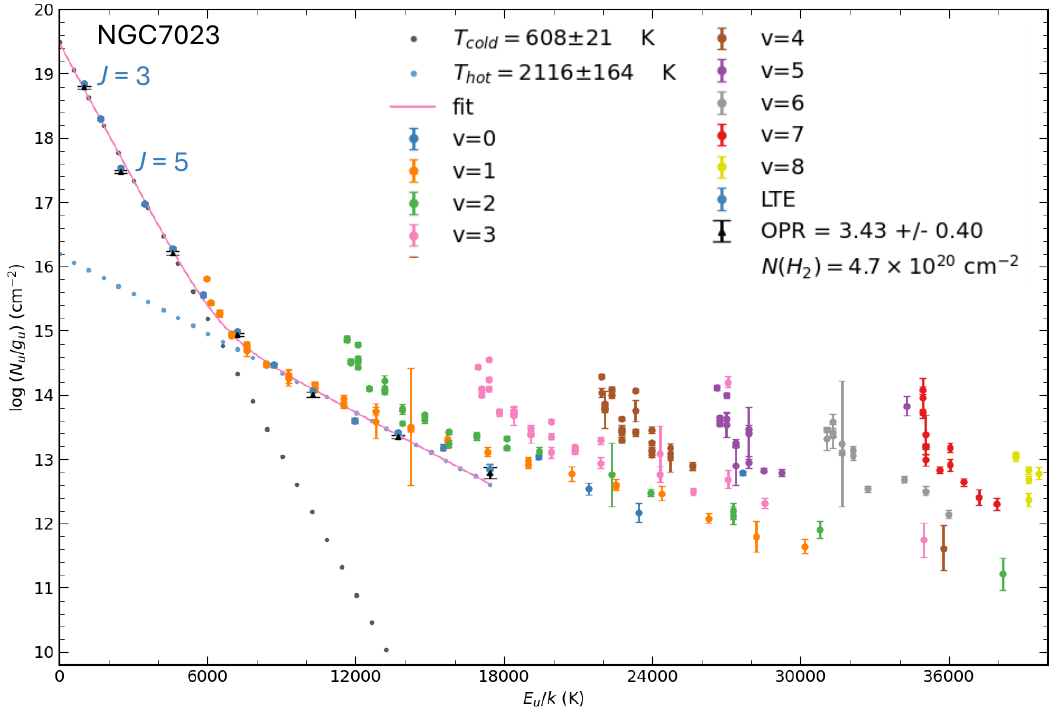}
		\includegraphics[width=\linewidth]{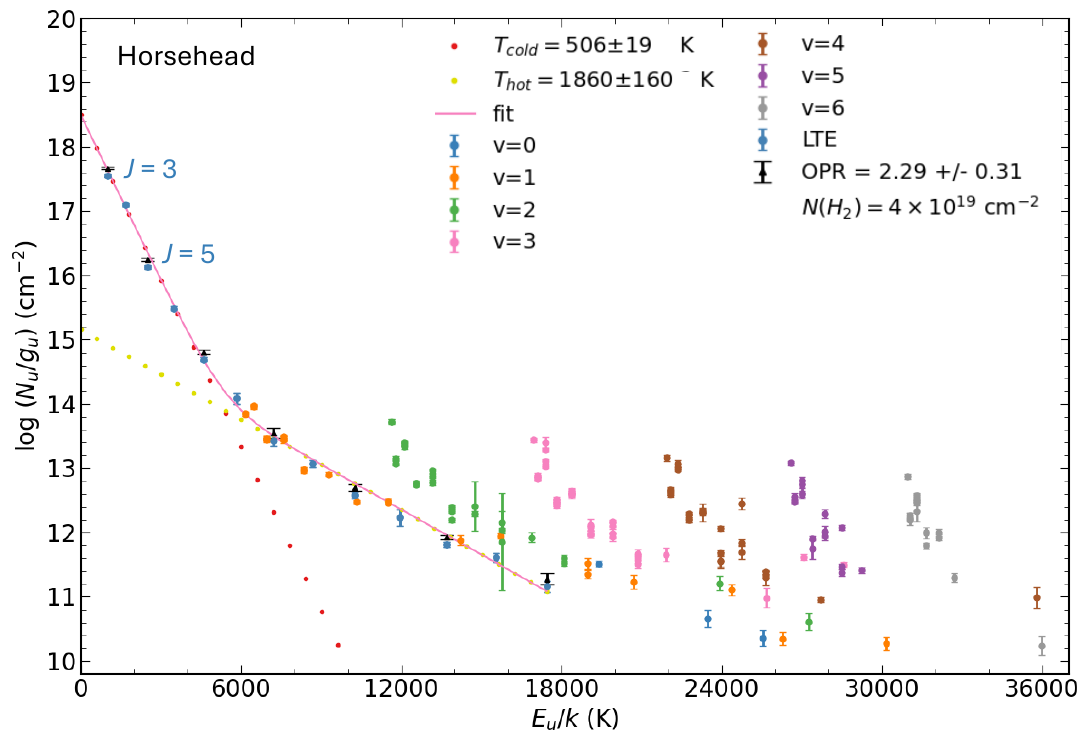}
		\caption{Excitation diagram with all the H$_2$ lines detected in the three PDRs: (top panel) Orion Bar, (middle panel) NGC7023, (bottom panel) Horsehead Nebula. The computed column density corresponds to $N({\rm H_2}) = N_{\rm cold}({\rm H_2}) + N_{\rm hot}({\rm H_2})$. The upper-level column densities are corrected for extinction using $A_V = 5.2$ for the Orion Bar and $A_V = 2.7$ for NGC7023, following Sect. \ref{sect:extinction}.}
		
		\label{fig:diag_rot}
	\end{figure}

	Figure \ref{fig:diag_rot} displays the excitation diagram of the brightest and/or the less extincted DF of each of the three PDRs. They were made using the \texttt{PhotoDissociation Region Toolbox Python module}\footnote{\url{https://github.com/mpound/pdrtpy}} \cite[pdrtpy,][]{Kaufman_2006,Pound_2008,Pound_2011,Pound_2023}. The intensities with and without the extinction correction applied ($A_V = 5.2$ for the Orion Bar and $A_V = 2.7$ for NGC7023, following Sect. \ref{sect:extinction}) for the lines used for the fitting procedure are presented in Table \ref{tab:intensity_h2}. Correcting the intensity with the value of $A_V$ derived from the S(3) line instead of NIR lines changes the derived column density by a factor of less than 2, and temperatures are only marginally affected. The three diagrams exhibit at least two components. The cold component corresponds to thermalized levels of \molh and can serve as a good measure of gas temperature. The hot component corresponds to the FUV-pumped levels.  Table \ref{tab:carac_PDRs} summarizes the main parameters derived from \molh lines.

	The cold component is very similar in all regions, around $T_{\rm cold} \sim 500$~K. The temperature is a bit lower in the Horsehead with $T_{\rm cold} = 506 \pm 19$~K, whereas NGC7023 and the Orion Bar are around $T_{\rm cold}(\text{Orion Bar}) = 580 \pm 16$~K and $T_{\rm cold}(\text{NGC7023}) = 608 \pm 21$~K, respectively. Given the different incident UV field intensities in these PDRs, this result may initially seem surprising. However, the incident UV field intensity is often measured at the IF. \molh emission traces the DF, which is located behind the atomic region, which, as is the case with the Orion Bar, can be quite large. To compare excitation temperatures, it is necessary to estimate the intensity of the UV field at the DF. In the Orion Bar, if we assume that the size of the atomic region is 10" (the distance to the first DF which has a similar gas temperature) and that the density inside this region is $n_{\rm H} = 5 \times 10^4 - 10^5$~cm$^{-3}$ \citep{elyajouri_2024}, then considering $R_V = 5.5$, we estimate a visual extinction at the DF around $A_V \sim 1-2$, which is in agreement with PDR models (Meshaka et al. in prep.). Considering the $R_V$-parametrized extinction curve from \cite{gordon2023} for $R_V = 5.5$, we then estimate that the UV field intensity is extincted by a factor of 5 to 15. Hence, if we consider $G_0(\rm IF) = 2 \times 10^4$ at the IF, then we expect that the UV field intensity at the DF3 to be around $G_0(\rm DF3) \sim 1300-5000$.  Interestingly, using an extinction curve evaluated at $R_V = 3.1$, representative of a diffuse medium, we find $G_0 (\rm DF3) < 1000$, which cannot explain the high temperature and high FUV-pumping excitation. This indicates that nanograins must be depleted in the atomic region of the Orion Bar, which flattens the UV portion of the extinction curve. In NGC7023, the UV field is also extincted by the dust and gas in front of the DF. However, even if the size of the "atomic" region of NGC7023 (the distance between the DF and the star) is significantly larger than the atomic region in the Orion Bar ($\sim 0.08$~pc versus $\sim 0.02$~pc for the Orion Bar), the density is lower ($n_{\rm H} \sim 10^3$~cm$^{-3}$). Hence, the UV field is not as extincted as in the Orion Bar. \cite{Pilleri12} estimated an extinction of $A_V = 1.5$ towards the star by comparing the \textit{International Ultraviolet Explorer} spectrum measured of the star and its Kurucz spectrum. If we assume that the same extinction factor applies between the star and the PDR, this leads to $G_0 $(DF) $= 2600$.
	In the DFs of the Horsehead Nebula, we expect the UV field to be very close to the one estimated from the projected distance as the atomic region is very small \citep[less than $5\times 10^{-4}$~pc,][]{abergel_jwst_2024}. Hence, we find that the intensity of the UV field at the DF is very similar in NGC7023 and the Orion Bar but significantly lower in the Horsehead Nebula. This agrees with the derived temperatures for the three PDRs: the Orion Bar and NGC7023 are similar, while the temperature for the Horsehead Nebula is lower. 
	
	For a regime where $G_0(\text{IF})/n_{\rm H} > 0.02$, PDR models predict that the position of the H/H$_2$ transition is settled mainly by dust extinction \citep[][]{Burton90}. The \molh DF is located at a depth in the cloud where the incident UV radiation field has been attenuated by dust extinction to such an extent that \molh formation on grains can counterbalance \molh photodissociation such that $2n($H$_2)/n$(H)$=1$. Therefore, when $G_0$ increases, in PDRs with similar gas density, the H/H$_2$ transition moves deeper into the PDR at a position with similar local UV intensity and gas temperature (see Fig. \ref{fig:temp_h_h2_GOn}). Finding similar physical conditions at the DF in the Orion Bar ($G_0(\text{IF})/n_{\rm H} \sim 0.1$) and NGC7023 ($G_0(\text{IF})/n_{\rm H} \sim 0.03$) is thus predicted by PDR models (see Fig. \ref{fig:temp_h_h2},\ref{fig:GDF_h_h2}). Following \cite{Sternberg2014} and \cite{Tielens:book2}, the extinction at 1000 $\AA$ at the DF where $x($H$_2) = 1/4$ can be written as
	\begin{equation}
		\tau(1000 \AA) = \frac{3}{4}\ln\left((\alpha G(1000 \AA))^{4/3}+1\right)
	\end{equation}
	with
	\begin{align}
		\alpha G =& G_0 \left(\frac{\sigma_d}{1.9 \times 10^{-21} \text{ cm$^{2}$ H-nuclei$^{-1}$}} \right)\left(\frac{10^2 \text{ cm}^{-3}}{n_{\rm H}} \right) \\
		& \times \left( \frac{3 \times 10^{-17} \text{ cm$^3$ s$^{-1}$}}{R_f} \right)\\
		& \times \left(1 + 8.9 \left( \frac{\sigma_d}{1.9 \times 10^{-21} \text{ cm$^{2}$ H-nuclei$^{-1}$}} \right)  \right)^{-0.37},
	\end{align}
	where  $\sigma_d$ is the UV dust grain absorption cross section, $n_{\rm H}$ is the proton density and $R_f = 3 \times 10^{-17}$~cm$^3$~s$^{-1}$ is the \molh formation rate. In the Orion Bar, $\sigma_d(1000\AA) \sim 6 \times 10^{-22}$~cm$^{2}$~H-nuclei$^{-1}$, hence, $\alpha G(1000 \AA) \sim 20G_0(\text{IF})/n$. In this region, $G_0(\text{IF})/n_{\rm H} \sim 0.1-1$, so we find $\tau(1000 \AA) \sim 1-3$. The mean extinction in the UV is around $\tau_{\rm UV} \sim 0.75 \tau(1000 \AA) \sim 0.8-2.3$. Finally, we find $G_0$(DF) $\sim 2700-9000$, which is consistent with a similar value of $G_0$ in the DF of NGC7023 and the Orion Bar.
	On the contrary, for a regime where $G_0(\text{IF})/n_{\rm H} < 0.02$ (in the Horsehead $G_0(\text{IF})/n_{\rm H} \sim 0.005$), H$_2$ self-shielding is becoming dominant and PDR models do not predict high gas temperature at the H/H$_2$ transition (see Fig. \ref{fig:temp_h_h2}).  Hence, the gas temperature observed at the DFs of the Horsehead Nebula cannot be explained by stationary models considering the low irradiation field \citep{Zannese_2025}. Imaging of the Horsehead Nebula revealed a regular pattern of “finger-like” structures resembling small-scale cometary globules observed in photoevaporating molecular clouds \citep{Bertoldi_1990,Lefloch_1994} and the ionization and DFs are not clearly spatially resolved and may even be merged. As discussed in \cite{Zannese_2025}, this suggests that dynamical effects are significant and may increase the gas temperature at the very edge, where there is mixing between the cold neutral medium and the more diffuse neutral and ionized medium. In addition, the ratio of the total luminosity of \molh lines over AIBs emission, which approximates the fraction of energy deposited in \molh excitation, is between 5 to 10 times higher in the Horsehead Nebula than in the Orion Bar and NGC7023. This could be explained by photoevaporation in the Horsehead Nebula, allowing a larger column density of \molh to be exposed to UV light.\\

	The hot component, linked to FUV-pumping, is also similar between the three PDRs, around $T_{\rm hot} \sim 2000$~K. We observe a trend where this component increases with incident $G_0$. We find $T_{\rm hot}(\rm Horsehead) \sim 1800$~K, $T_{\rm hot}(\rm NGC7023) \sim 2100$~K and $T_{\rm hot}(\rm Orion~Bar) \sim 2500$~K. Determining whether this component is sensitive to the absolute intensity of the incident UV field or to the spectral energy distribution, i.e. shape, associated with O versus B star requires detailed modeling, which is beyond the scope of this paper.
	The breakpoint in the slope between the cold and the hot components differs for each PDR. In the Orion Bar and NGC7023, the \molh distribution deviates from LTE around $J=8-9$; and in the Horsehead, around $J=7-8$. As the observed temperatures are very similar between the three PDRs, the difference in this break point is linked to the local gas density. Table \ref{tab:crit_dens} presents the critical densities of the first detected \molh levels. To first order, the critical density of the levels at the transition between LTE and FUV-pumping excitation should be roughly equal to the local gas density. Hence, we find $n(\text{Orion Bar/NGC7023})~\sim~(1-2)~\times~10^5$~cm$^{-3}$, and $n(\text{Horsehead})~\sim~5\times~10^4$~cm$^{-3}$. After filament size, this is the second piece of evidence that the Horsehead is the least dense PDR, and that the Orion Bar and NGC7023 seem to have very similar densities.
	
	\begin{table}[!h]
		\centering
		\caption{Critical densities of \molh rotational levels at $T = 500$~K for collisions with \molh, H, and both \molh and H.}
		\begin{tabular}{c|c|c|c}
			Level   & $n(\rm H_2)_{\rm cr}$\tablefootmark{a} (cm$^{-3}$) & $n(\rm H)_{\rm cr}$\tablefootmark{b}  (cm$^{-3}$) & $n_{\rm tot_{\rm cr}}$\tablefootmark{c}  (cm$^{-3}$)\\\hline\hline
			$v=0$ $J=3$ & $1.9 \times 10^2$ & $1.4 \times 10^3$ & $1.6 \times 10^2$ \\\hline
			$v=0$ $J=4$ & $2.0 \times 10^3$ & $5.8 \times 10^3$ & $1.5 \times 10^3$\\\hline
			$v=0$ $J=5$ & $1.6 \times 10^4$ & $1.5 \times 10^4$ & $7.8 \times 10^3$ \\\hline
			$v=0$ $J=6$ & $8.6 \times 10^4$ &  $3.2 \times 10^4$ & $2.4 \times 10^4$ \\\hline
			$v=0$ $J=7$ & $ 3.4 \times 10^5$ & $6.4 \times 10^4$ & $5.4 \times 10^4$ \\\hline
			$v=0$ $J=8$ & $1.1 \times 10^6$ & $1.2 \times 10^5$ & $1.1 \times 10^5$ \\\hline
			$v=0$ $J=9$ & $3.2 \times 10^6$ & $1.9 \times 10^5$ & $1.8 \times 10^5$ 
		\end{tabular}
		\tablefoot{The critical density of a level $i$ is calculated as $n_{{\rm cr}_{i}} = \frac{\sum_j A_{ij}}{\sum_j k_{ij}}$ where $A_{ij}$ is the Einstein coefficient from the level $i$ to a level $j$ and $k_{ij}$ is the collisional de-excitation rate from a level $i$ to a level $j$. Collisional data from \tablefoottext{a}{\cite{Flower_1998,Flower_1999}},\tablefoottext{b}{\cite{Lique_2015,Bossion_2018}},\tablefoottext{c}{Considering $n_{\rm tot} = n({\rm H}) + n({\rm H_2}) = 2n({\rm H}) = 2n({\rm H_2})$} (DF). Collisions with electrons, considering an abundance of $x(e^-) = 10^{-4}$, are negligeable compared to collisions with H and H$_2$.} 
		
		\label{tab:crit_dens}
	\end{table}

	In the Orion Bar and NGC7023, the $v=0$ and $v=1$ ladders exhibit similar behavior. This can be explained by the fact that, in these PDRs, the density is high enough to thermalize the first levels (low-$J$) of the $v=1$ ladder ($n_{\rm cr}$($v=1$, $J=3$) $\sim 10^5$~cm$^{-3}$). This is not the case for the Horsehead Nebula, where the $v=1$ ladder seems to follow the hot component of the rotational levels, which is still in agreement with the Horsehead Nebula being less dense. The $ v=2, 3, 4, 5,$ and $6$ form separate ladders composed of two components with rotational temperatures similar to those for $v=0$. The higher $J$ levels converges with $v = 0$ and $v = 1$ ladder. The population distribution of the lower $J$ levels in high $v$ states is characteristic of fluorescently excited \molh. Indeed, this mechanism populates the very low-$J$ of vibrationally excited levels. This was modeled with the Meudon PDR code with parameters representative of the Orion Bar and the Horsehead in \cite{Maillard_2023} (see their Figs. 8.19 and 8.20, each peak in the fluorescent excitation corresponds to the low-$J$ levels) and Piluso et al. submitted (see their Fig. 4). The levels aligned with the hot component are then populated by UV pumping followed by an IR cascade.

	\subsection{Ortho-to-para ratio}
	
	For both NGC7023 and the Orion Bar, we find an \molh OPR for rotational lines around 3 at the dissociation fronts, which is expected as the gas temperature is a lot higher than the equilibrium temperature $T_{\rm eq} \sim 200$~K. Hence, contrary to the Horsehead Nebula, no presence of out-of-equilibrium mechanisms such as the ortho-para conversion on grains \citep{bron_efficient_2016} is observed, nor the advection and mixing of colder \molh in the warm region \citep{gorti_photoevaporation_2002, storzer_nonequilibrium_1998}. 
	
	\cite{Zannese_2025} showed that in the Horsehead Nebula this ratio is lower in the vibrational levels, around OPR$_{\rm rovib}~\sim~\sqrt{\text{OPR}_{\rm rot}}$, which is explained by a preferential self-shielding of ortho levels compared to para levels. Such self-shielding favors the pumping of para levels and thus the para vibrational transitions \citep{sternberg}. This is somewhat less visible in the Orion Bar and NGC7023. Indeed, the OPR remains 3 for all levels, even at high $J$ or high $v$ for the Orion Bar (see Table \ref{tab:OPR}). The OPR appears to be slightly lower for more excited levels in NGC7023 but remains higher than $\sqrt{\text{OPR}_{\rm rot}}$, except for high vibrational levels $v \geq 2$. The fact that, in these denser PDRs, we do not retrieve a signature of UV pumping in the OPR value might reveal the importance of collisions, enabling collisional flipping of the spin, within the vibrational levels and "thermalize" them to an OPR closer to 3.
	
	\begin{table}[!h]
		\centering
		
		\caption{Variation of the Ortho-to-Para ratio considering \molh levels.}
		\begin{tabular}{c|c|c|c}
			Region  & Orion Bar & NGC7023 & Horsehead \\
			\hline \hline
			OPR $v=0$ $J<7$ & 3.6 $\pm$ 0.7 & 3.6 $\pm$ 0.2 & 2.4 $\pm$ 0.8\\\hline
			OPR $v=0$ $J>10$ & 3.1 $\pm$ 0.5 & 3.0 $\pm$ 0.7 & 2.1 $\pm$ 0.3 \\\hline
			OPR $v=1$ $J=1-3$\tablefootmark{a} & $2.6 - 4.0$ & $2.0 - 3.0$ & $1.4 - 1.8$\\\hline
			OPR $v=2$ $J=0-2$\tablefootmark{a} & $2.3 - 3.5$ & $1.2 - 2.0$ & $0.9 - 1.4$\\\hline
			OPR $v=3$ $J=0-2$\tablefootmark{a} & $1.5 - 3.0$ & $1.0 - 1.4$ & $0.8 - 1.4$\\
			
		\end{tabular}
		\tablefoot{\tablefoottext{a}{The range of OPR values has been evaluated using several lines originating from the same level.}}
		\label{tab:OPR}
	\end{table}

	\section{Physical conditions derived from \molh}
	\label{sect:physcial_conditions}
	\subsection{Variation of the gas temperature}

	\begin{figure}
		\centering
		\includegraphics[width=0.9\linewidth]{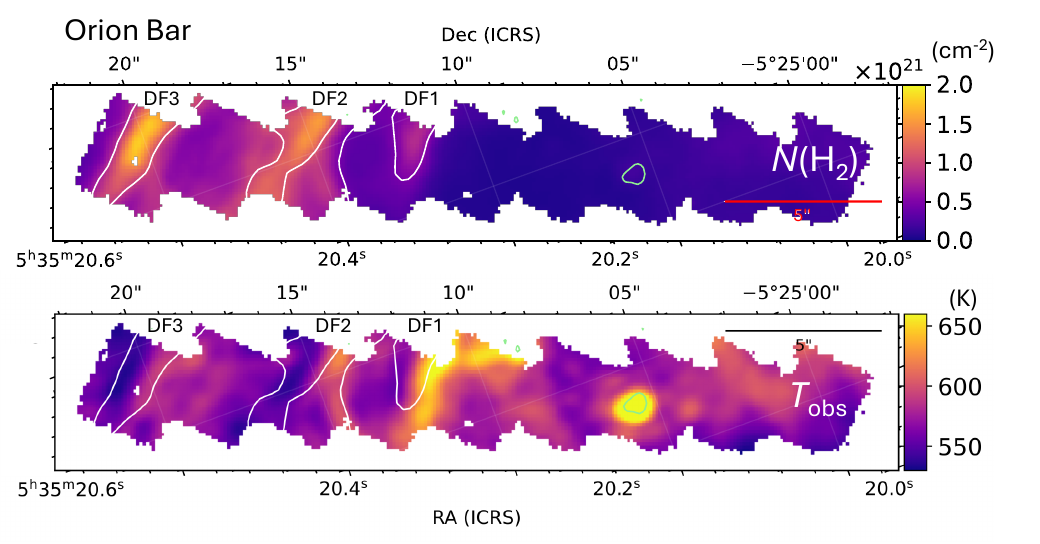}
		\includegraphics[width=0.9\linewidth]{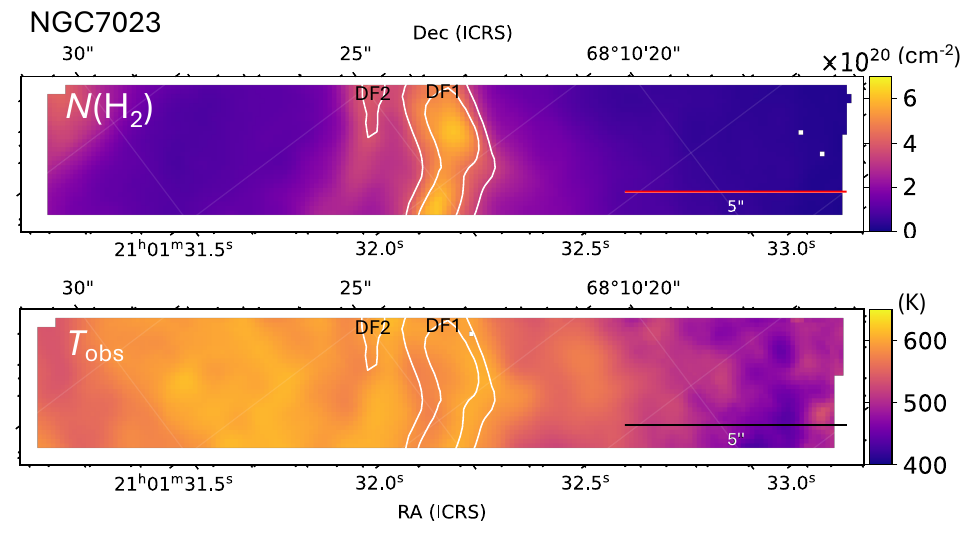}
		
		\includegraphics[width=0.9\linewidth]{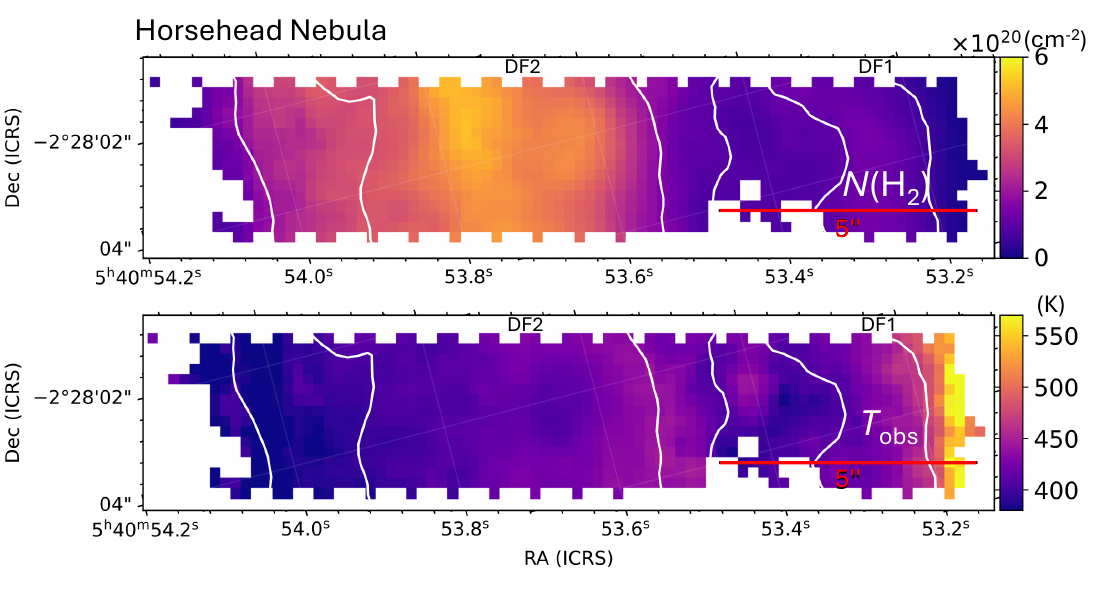}
		\caption{Maps of the observed column density and temperature across the Orion Bar (top panel), NGC7023 (middle panel), and Horsehead Nebula (bottom panel). White contours are from the 0–0 S(1) line emission. Green contours focus on the protoplanetary disk d203-506 and are from 1–0 S(1).}
		\label{fig:temp_N_AV_map}
	\end{figure}
	\begin{figure*}
		\centering
		\includegraphics[width=0.85\linewidth]{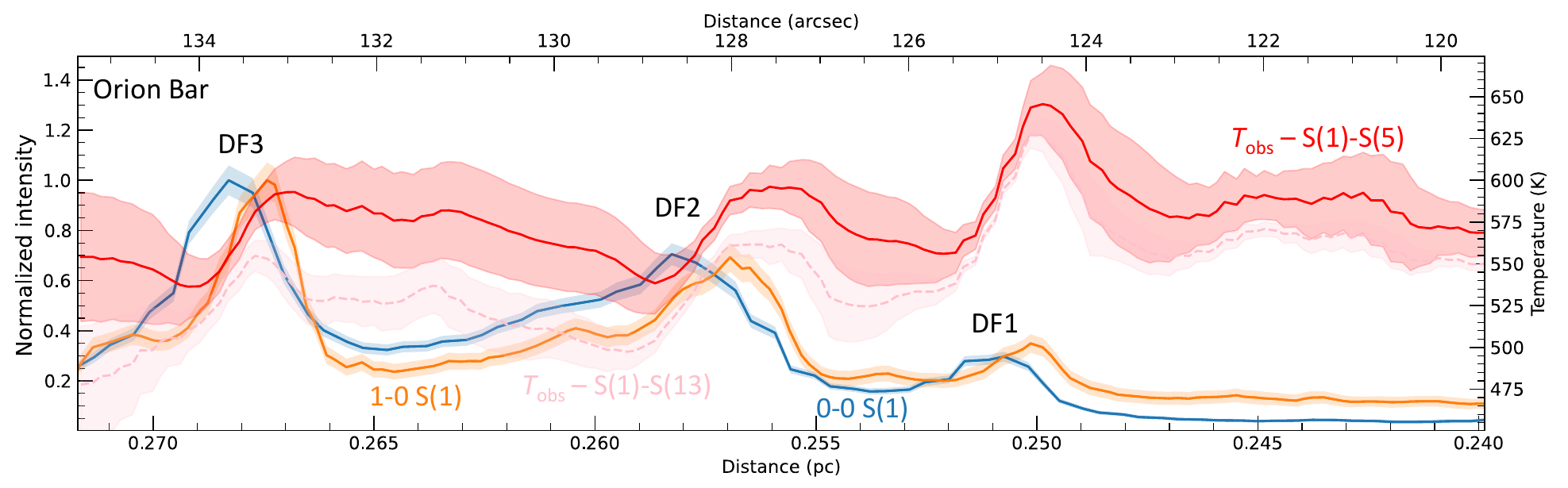}    
		\includegraphics[width=0.85\linewidth]{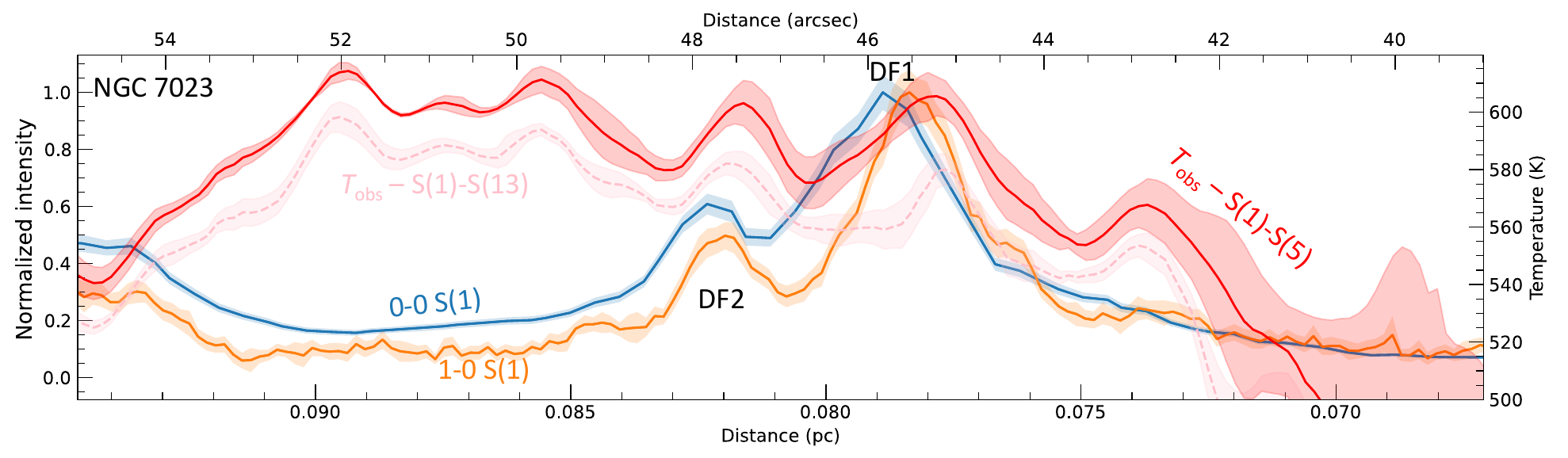}
		
		\includegraphics[width=0.84\linewidth]{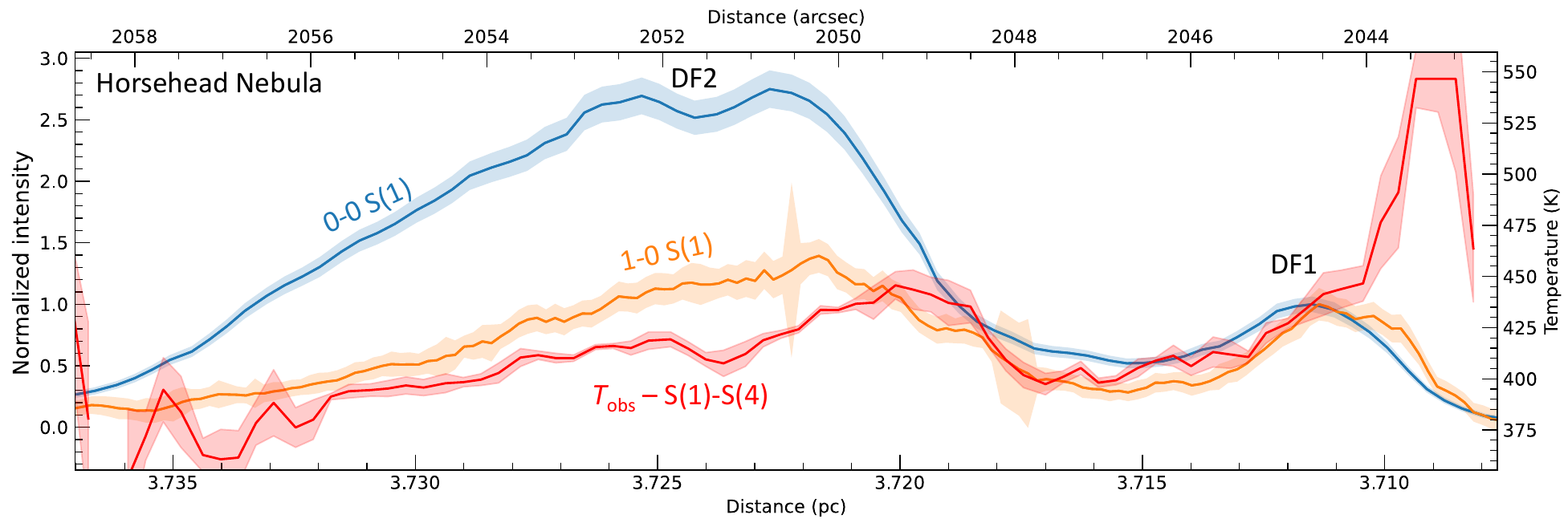}
		\caption{Gas temperature profiles derived from a single-component fit of the first five (resp. four for the Horsehead Nebula) observed lines (S(1)- S(5), or S(1) - S(4)) without the S(3), compared to the \molh 0–0 S(1) and 1–0 S(1) line emission profile in (top panel) the Orion Bar, (middle panel) NGC7023, and (bottom panel) the Horsehead Nebula as a function of the distance to the illuminating star (pc and arcsec). The dashed pink line corresponds to the gas temperature derived from the cold component obtained by fitting the first observed \molh (S(1)–S(15) or S(13)) lines with two components. A slight decrease in temperature is observed in the different \molh filaments.}
		\label{fig:H2_Temp_cut}
	\end{figure*}

	Using the pdrtpy module, we derived maps of temperature and column density by fitting the $0-0$ S(1) -- S(5) (which show a straight line shape in the excitation diagrams, implying a single-excitation temperature close to $T_{\rm gas}$) lines ignoring the 0-0 S(3) line (see Sect. \ref{sect:extinction} for details). These levels are thermalized, and the resulting temperature is the gas temperature. We fit these lines with a single component, keeping the OPR as a free parameter. Fig.~\ref{fig:temp_N_AV_map} presents the temperature and column density maps in the Orion Bar,  NGC7023, and the Horsehead Nebula. 
	The OPR map is not shown, as it remains $\sim 3$ across Orion Bar and NGC7023 regions \citep[for the Horsehead nebula, see Fig. 8 of][]{Zannese_2025}. The column density maps mimic the structure observed in the \molh line maps (see Fig.~\ref{fig:H2_maps}). The column density of \molh peaks at each DF. We obtain very similar maps using either value of $A_V$ (from the S(3) line or NIR lines). The temperature maps are almost identical, and the column density maps are slightly changed (for instance, in the Orion Bar, DF1 is brighter when using the value of $A_V$(S(3)) but the overall variations are the same.
	
	Interestingly, the temperature maps show little spatial variation across an individual PDR. These results are in agreement with what was found for the Horsehead Nebula. 
	In the Orion Bar, the temperature appears to be highest at the edge of the DFs, similarly to the Horsehead Nebula. In NGC7023, this result is much less clear, as the temperature seems very homogeneous across the FOV. Fig.~\ref{fig:H2_Temp_cut} shows the observed temperature profile along the cuts presented in Fig.~\ref{fig:FOVbis}.
	Similar to the Horsehead Nebula, we observe a decrease in temperature in each DF. Fig. \ref{fig:cutfusion} in the Appendix shows the same profile when using data fusion. Thanks to the increased spatial resolution of the rotational lines, we resolve this temperature decrease more clearly in the DFs of NGC7023. These decreases are consistent with the extinction of the UV field within the PDR. However, these small variations in temperature, similar to those found in the Horsehead Nebula, reveal that the illuminated matter dominates the emission of \molh (see schematic view of the three PDRs in Fig. \ref{fig:geometry}). Indeed, in both the Orion Bar and NGC7023, we observe high temperatures after the fronts, probably originating from \molh in the foreground or background. Hence, \molh emission is not a good tracer of the cooler gas within the PDR. Similarly, the temperature probed ahead of the DFs, toward the atomic region, is tracing the  the illuminated \molh layer in the background rather than the less abundant \molh in the atomic region. 
	
	In NGC7023, the temperature is similar at each front ($T \sim 600$~K). In contrast, in the Orion Bar, the temperature at the DFs decreases with distance from the star. Indeed $T_{\rm DF1} \sim 645 \pm 25$~K, $T_{\rm DF2} \sim 595 \pm 15$~K and $T_{\rm DF3} \sim 590 \pm 15$~K. Contrary to NGC7023, the fronts are farther apart from each other ($d \sim 0.01$~pc $\sim 2000$~au versus $d\sim 0.003$~pc $\sim 600$~au in NGC7023), so this is not that surprising. In addition, the extinction map shows that DF1 is located further away from the observer than DF2 and DF3, so the projected distances are most certainly different from the real distance to the star. The fact that we retrieve a higher temperature in DF1 might mean that it is closer to the irradiating star.

	\subsection{Estimation of gas density}
	
	\begin{figure}
		\centering
		\includegraphics[width=\linewidth]{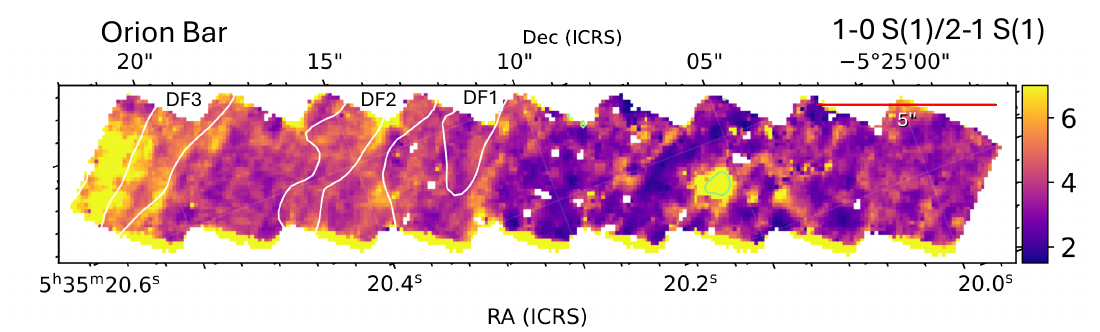}
		\includegraphics[width=\linewidth]{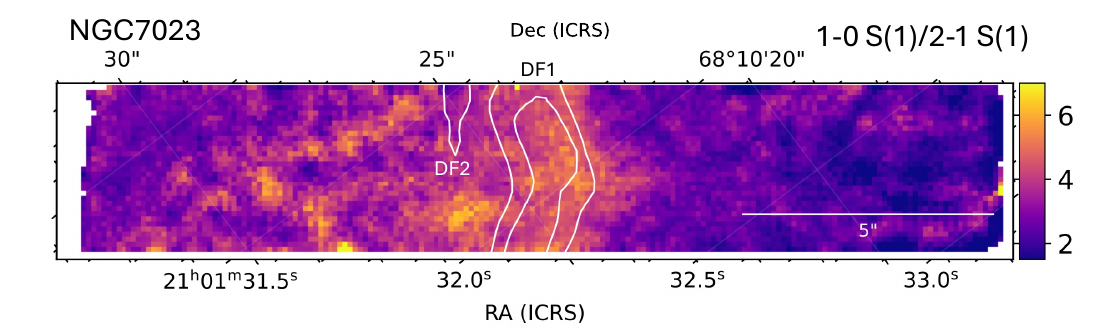}
		\includegraphics[width=\linewidth]{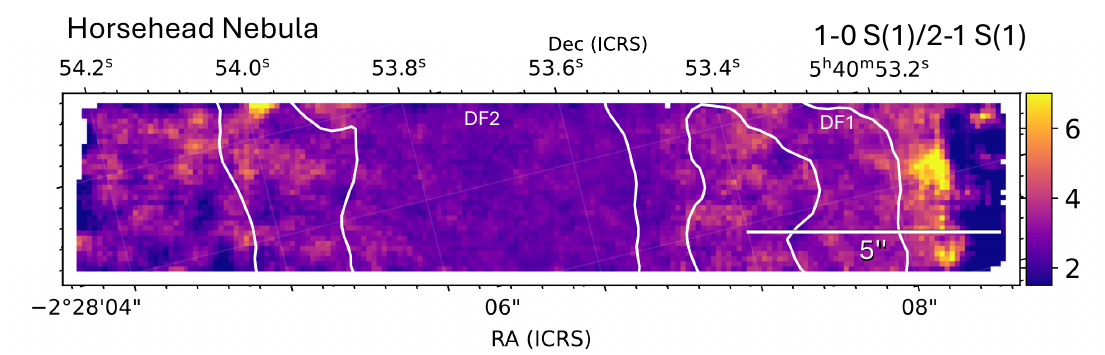}
		
		\caption{$1-0$ S(1) / $2-1$ S(1) maps as a tracer of gas density. White contours are from the 0–0 S(1) line emission. Green contours focus on the protoplanetary disk d203-506 and are from 1–0 S(1).}
		\label{fig:ratio}
	\end{figure}
	
	To first order, we can use the measured \molh column density to estimate upper limits for the gas density.  In the Orion Bar, at the peak of DF3\footnote{Here, we take the maximum column density and not the averaged value over the filament,  as done in Fig. \ref{fig:diag_rot}.}, the column density is around $N($H$_2) \sim 1.8 \times 10^{21}$~cm$^{-2}$. We assume that the size of \molh emission in the line of sight is at least the width of the H$_2$ emission in the plane of the sky, i.e., around 1". With this width, we can estimate the upper limit of \molh density as $n($H$_2) < 3 \times 10^5$~cm$^{-3}$. In NGC7023, the column density of \molh at the peak of DF1 is around $N($H$_2) \sim 6 \times 10^{20}$~cm$^{-2}$. The size of the \molh emission is similar to that in the Orion Bar, around 1". The upper limit of \molh density in NGC7023 is thus $n($H$_2) <  10^5$~cm$^{-3}$. Using the same method for the Horsehead Nebula, we find $N($H$_2) \sim 1.3 \times 10^{20}$~cm$^{-2}$ at the peak of DF1, and we derive $n($H$_2) <  1.5 \times 10^4$~cm$^{-3}$ at the DF. The total density at the DF can be about twice as high due to the presence of atomic hydrogen. To evaluate the density difference between the PDRs, we can also use the $1-0$ S(1)/$2-1$ S(1) line ratio, which is a tracer of density under dense, highly irradiated conditions. Under these conditions, collisional excitation of the \molh $v=1$, $J=3$ level becomes competitive when the density increases above $10^5$ cm$^{-3}$. The $1-0$S(1)/$2-1$S(1)  line ratio is thus expected to increase from a pure radiative cascade value (about 2) to a collisional excitation value (of the order of 10). In the template DFs, we find $5.1 \pm 0.5$ in the Orion Bar, $4.4 \pm 0.3$ in NGC7023, and $3.8 \pm 0.4$ in the Horsehead Nebula. 
	
	Throughout this study, we have presented various diagnostics for the gas density in these PDRs (increasing \molh emission size from the Orion Bar to the Horsehead Nebula; see Sect. \ref{sect:spat_morph}, positions of the breakpoint between thermalized and FUV-pumped levels excitation diagrams in Sect. \ref{sect:diag_rot}, $1-0$ S(1)/$2-1$ S(1) line ratio, direct estimation from the column density). All these diagnostics point to the same result: we find increasing density from the Horsehead Nebula to the Orion Bar, with NGC7023 in between. This result is consistent with the difference in IR line richness among the spectra of the three objects. Indeed, many molecular lines are detected in the Orion Bar \citep{Joblin_2018,Peeters_2024,van_de_putte_2024,Zannese_2025}, whereas almost only \molh is detected in the Horsehead Nebula \citep{Misselt_2025,Lis_2026} even though the SNR is sufficient to detect molecular lines with the same strength as those in the Orion Bar. NGC7023 seems to be an intermediate between the two other objects, where some molecular lines requiring high densities are detected \citep[e.g. CH$^+$,][]{Joblin_2018,Misselt_2025} but with lower intensities than observed in the Orion Bar. This density difference between these PDRs is also reproduced with precise modeling using the Meudon PDR Code \citep{Le_Petit_2006}. Piluso et al., submitted find a best fit model for the Horsehead Nebula at thermal pressure of $P_{\rm th}/k_B = (5.4 - 7.2)\times 10^6$~K~cm$^{-3}$ ($n_{\rm H} \sim 10^{4}$~cm$^{-3}$) and for the Orion Bar at  $P_{\rm th}/k_B = (4.9 - 5.6)\times 10^7$~K~cm$^{-3}$ ($n_{\rm H} \sim 10^{5}$~cm$^{-3}$).
	
	The $1-0$S(1)/$2-1$S(1) ratio can also be used to evaluate density variations within the PDR FOV. Fig.~\ref{fig:ratio} presents this ratio in the FOV for the Orion Bar, NGC7023, and the Horsehead Nebula. For the Horsehead Nebula, we recover the results from \cite{Zannese_2025} and \cite{elyajouri_2025} with a very steep PDR front and high density near the edge of the PDR. The ratio reaches 6 at the edge and then decreases to 2, a value characteristic of a pure radiative cascade, around the second DF, located in the center of the FOV. This variation suggests that DF1 is denser than DF2. The high density and, therefore, overpressure near the edge of the Horsehead Nebula are compatible with important dynamical effects in this region \citep[for a more detailed discussion, see][]{Zannese_2025}. In NGC7023, the ratio is only well measured near DF1 and DF2. We observe an increase in density between the atomic and dissociation regions, as expected for an isobaric PDR, and no strong variations between the two fronts. In the Orion Bar, the density appears rather constant in the dissociation-front region and increases after DF3, with the ratio rising from around 3 to around 8. This increase in density is consistent with the presence of high density (up to a few 10$^6$~cm$^{-3}$), cooler molecular gas structures \citep[Zannese et al. submitted]{Lis03,Goicoechea_2016}. However, this rise is at the very edge of the FOV, so strong conclusions are difficult to draw.

	\section{Conclusions}
	\label{conclusions}
	In this work, we investigated and compared the emission spatial morphology and excitation of \molh in three emblematic PDRs: the Orion Bar, NGC7023, and the Horsehead Nebula. We studied in detail the different excitation processes of \molh: fluorescence and IR cascade after UV pumping, and collisions.  
	Leveraging the high sensitivity and high spatial resolution, we have been able to obtain several constraints on \molh excitation:
	\begin{itemize}
		
		\item In the three PDRs, the morphology of \molh emission follows a filamentary structure with small scales (1-1.5"). We recover a spatial shift between different \molh lines, revealing different excitation processes (UV pumping vs collisions) and temperature gradient ($0-0$ S(1) peaks deeper in the PDR than $0-0$ S(3)). 
		\item The analysis of \molh excitation diagrams reveals the different excitation mechanisms: collisions for the lowest rotational levels, FUV-pumping for rovibrational levels and more excited rotational levels. We also find that the vibrational levels with low$-J$ are efficiently populated by fluorescence directly. 
		\item In the Orion Bar, we find that the OPR remains 3 for all \molh levels, contrary to what is found for the Horsehead Nebula and NGC7023, where the OPR of rovibrational levels is closer to $\sqrt{\rm OPR_{rot}}$ due to  excitation via FUV pumping. This might reveal the importance of collisions even in vibrational levels in the Orion Bar, the densest PDR of the three, which allows the "thermalization" of these levels.
		
	\end{itemize}
	
	The analysis of \molh excitation has allowed us to constrain several physical parameters (visual extinction, temperature, and density). First, using both the rovibrational levels of \molh and the $0-0$ S(3) line, we have been able to derive the extinction across the FOV, providing insight into the geometry of the PDRs and the composition of the grains. The two methods yield different estimates for $A_V$ and distinct spatial morphologies. This reveals that the extinction curve must be flatter than the one derived by \cite{gordon2023} with $R_V = 5.5$ and that it must be even flatter near the IF than deeper inside the PDR, compatible with a depletion of nanograins at the edge of the PDR.
	
	Then, the analysis of thermalized levels has allowed us to constrain the gas temperature throughout the PDRs: 
	\begin{itemize}
		\item From \molh rotational excitation, we derive very high temperatures in the three PDRs ($T_{\rm gas} \sim 500-600$~K). Interestingly, the temperatures are very similar in the Orion Bar and NGC7023, 
		a result that may initially seem surprising given their supposedly different FUV irradiation. This indicates that the atomic region in the Orion Bar is sufficiently wide and dense to efficiently attenuate the UV field at the DF to a value close to what is expected in NGC7023. 
		\item The high temperature found in the Horsehead Nebula cannot be explained by stationary models, considering the low-irradiation, and is probably due to important dynamical effects at the edge of the PDR, such as advection and mixing between the cold neutral medium and the more diffuse neutral and ionized medium \citep[for more details, see][]{Zannese_2025}.
		\item Similar to what was concluded for the Horsehead Nebula, we do not observe a large variation in temperature across the FOV in NGC7023 and the Orion Bar. This reveals that the thin, warm, illuminated layer ($\sim 2-3 \times 10^{-3}$~pc) dominates the \molh emission across the entire JWST FOV, due to the complex 3D geometry of the PDR. Thus, access to the cooler molecular gas is challenging via \molh emission, and its analysis must be complemented by other tracers of cool gas with ALMA at comparable angular resolutions \citep[e.g., CO,][]{Goicoechea_2016,hernandez-vera_extremely_2023}.
	\end{itemize}
	
	Finally, we have been able to derive several diagnostics of the gas density from the analysis of \molh emission. These diagnostics all show that the gas density increases from the Horsehead Nebula to the Orion Bar, with NGC7023 lying between the two. This result explains the clear difference in IR molecular lines richness, which is highly density-dependent, among these three PDRs. The different diagnostics are:
	
	\begin{itemize}
		\item The size of the \molh emission is larger in the Horsehead Nebula than in the Orion Bar and NGC7023, which shows that the Horsehead is the least dense PDR. This is because in a denser environment, the UV field is extincted over a shorter distance, and thus \molh emission is less extended.
		\item In the excitation diagram, the breakpoint between thermalized and FUV-pumped levels is linked to the gas density (as the gas temperature is very similar between the three PDRs). The critical density of the level coinciding with the breakpoint should be roughly the same as the gas density. We find that this breakpoint occurs at different levels across the three PDRs ($J=7-8$ for the Horsehead, $J=8-9$ for NGC7023 and the Orion Bar).
		\item The $1-0$S(1)/$2-1$S(1) ratio, which is a tracer of density, has an increasing value from the Horsehead Nebula ($3.8 \pm 0.4$) to NGC7023 ($4.4 \pm 0.3$) to the Orion Bar ($5.1 \pm 0.5$).
		\item Direct estimation of upper limits of the density using the column density of \molh also gives increasing values from the Horsehead Nebula ($n(\text{Horsehead}) < (1.5-3) \times 10^4$~cm$^{-3}$) to NGC7023 ($n(\text{NGC7023}) < (1-2) \times 10^5$~cm$^{-3}$) to the Orion Bar ($n(\text{Orion Bar})< (3-6) \times 10^5$~cm$^{-3}$).
	\end{itemize}
	
	In addition to estimating gas density for the three PDRs, we have also determined density variations within the PDRs using \molh analysis. We observe density variations across the PDR. In the Horsehead Nebula, we find a higher density at the edge of the PDR than deeper inside, which is compatible with strong dynamical effects. In the Orion Bar, we observe a density gradient at the edge of the FOV, after the DF3, consistent with the presence of high-density, cooler molecular gas structures \citep[Zannese et al. submitted]{Lis03,Goicoechea_2016}.
	Thanks to JWST's high sensitivity and spatial resolution, we can now probe these density variations at such small scales, for the first time. However, the complex 3D geometry of PDRs makes it challenging to analyze cooler \molh emission, as it is dominated by a thin, warm, illuminated layer across the entire JWST FOV. \molh emission provides strong constraints on physical conditions near the DF but must be combined with other tracers to probe the precise variations deeper inside the PDR.

	\begin{acknowledgements}
		
		This work is based on observations made with the NASA/ESA/CSA James Webb Space Telescope. The data were obtained from the Mikulski Archive for Space Telescopes at the Space Telescope Science Institute, which is operated by the Association of Universities for Research in Astronomy, Inc., under NASA contract NAS 5-03127 for JWST. These observations are associated with program \#1288 (DOI: 10.17909/pg4c-1737). Support for program \#1288 was provided by NASA through a grant from the Space Telescope Science Institute, which is operated by the Association of Universities for Research in Astronomy, Inc., under NASA contract NAS 5-03127, and the Canadian Space Agency (CSA, 22JWGO1-16). M.Z. and J.R.G. thank the Spanish MCINN for funding support under grant PID2023-146667NB-I00. M.Z. acknowledges the Juan de la Cierva Postdoctoral Fellow project JDC2024-054658-I, funded by MICIU/AEI/10.13039/501100011033 and by the ESF+. M.W.P. acknowledges support from NASA Astrophysics Data Analysis Program award \#80NSSC19K0573. E.P. and J.C. acknowledge support from the University of Western Ontario, the Institute for Earth and Space Exploration, the Canadian Space Agency (CSA, 22JWGO1-16), and the Natural Sciences and Engineering Research Council of Canada. C.B. is thankful for an appointment at NASA Ames Research Center through the San Jos\'{e} State University Research Foundation (80NSSC22M0107) and gratefully acknowledges support from the Internal Scientist Funding Model (ISFM) Laboratory Astrophysics Directed Work Package at NASA Ames. K.D.G and A.N-C are partially supported by NASA grant 80NSSC21K1294. K.M. is supported by JWST–NIRCam contract no. NAS5-02015 to the University of Arizona. T.O. acknowledges the support by the Japan Society for the Promotion of Science (JSPS) KAKENHI Grant Number JP24K07087.
		
	\end{acknowledgements}
	
	\bibliographystyle{aa} 
	\bibliography{mainbib} 

	\begin{appendix}
		\section{Estimation of the visual attenuation}

\begin{figure}[!h]
			\centering
			
			\includegraphics[width=\linewidth]{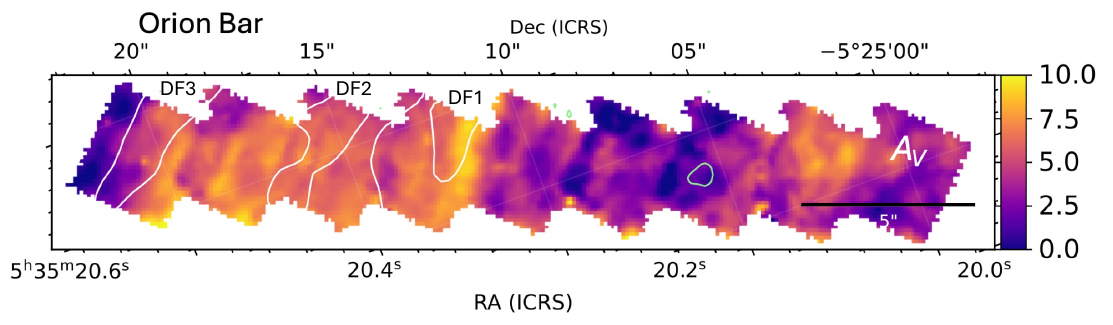}
			\includegraphics[width=\linewidth]{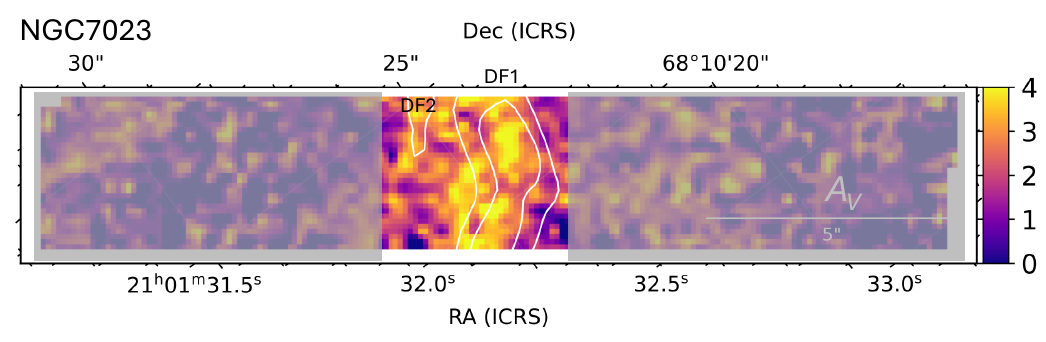}
			\caption{Maps of the visual attenuation in the line of sight for the Orion Bar (top) and NGC7023 (bottom). White contours are from the $0-0$ S(1) line emission. Green contours focus on the protoplanetary disk d203-506 and are from $1-0$ S(1). The grayed part is the region where the SNR of the \molh lines is very low, and thus the estimation of $A_V$ is uncertain. }
			\label{fig:AVmap}
		\end{figure}

\begin{figure}[!h]
			\centering
			
			\includegraphics[width=\linewidth]{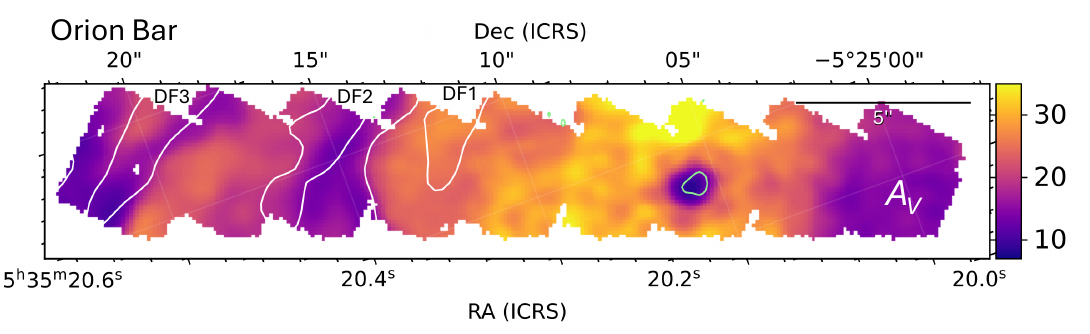}
			\includegraphics[width=\linewidth]{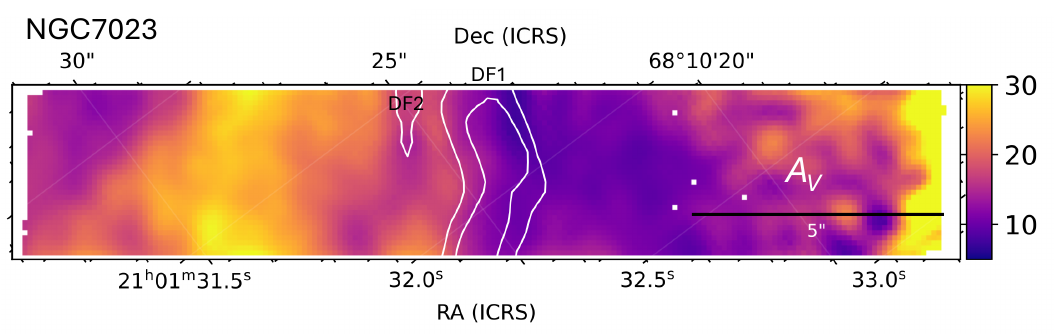}
			\caption{Maps of the visual attenuation in the line of sight for the Orion Bar (top) and NGC7023 (bottom), derived from the $0-0$ S(3) line. White contours are from the $0-0$ S(1) line emission. Green contours focus on the protoplanetary disk d203-506 and are from $1-0$ S(1). }
			\label{fig:AVmap_S3}
		\end{figure}

		In the study, we used the $R_V$-parametrized extinction curve from \cite{gordon2023} with $R_V = 5.5$. We evaluated the attenuation in the line of sight using two methods: (1) the NIR \molh line ratio coming from the same upper level and (2) the intensity of the $0-0$ S(3) line and considering an OPR of 3 (see Sect. \ref{sect:extinction} for more details). Fig.~\ref{fig:AVmap} (resp. Fig.~\ref{fig:AVmap_S3}) displays the spatial morphology of the attenuation using method (1) (resp. method (2)). The overall morphology of the attenuation differs markedly between the two methods, suggesting a flatter extinction curve in the near-IR than assumed in this study and a variation in its shape across the PDRs.
		
		\section{PDR models}
		\label{appendix:pdrmodels}
		To complement our observational study, we use the online grid of isochoric models of the Meudon PDR code\footnote{Grid of isochoric models from August 2024: \url{https://app.ism.obspm.fr/ismdb/}} \citep[][version 7.1]{le_petit_model_2006}. The code simulates the thermal and chemical structure of the gas in a self-consistent manner, assuming a 1D geometry and a stationary state in a plane-parallel irradiated gas and dust layer. The incident UV radiation field is that of an O5 star. The code includes progressive attenuation of the UV field due to grain and gas extinction. The extinction curve used is the mean Galactic extinction curve, parameterized by \cite{fitzpatrick_analysis_1988}.
		\begin{figure}
			\centering
			\includegraphics[width=\linewidth]{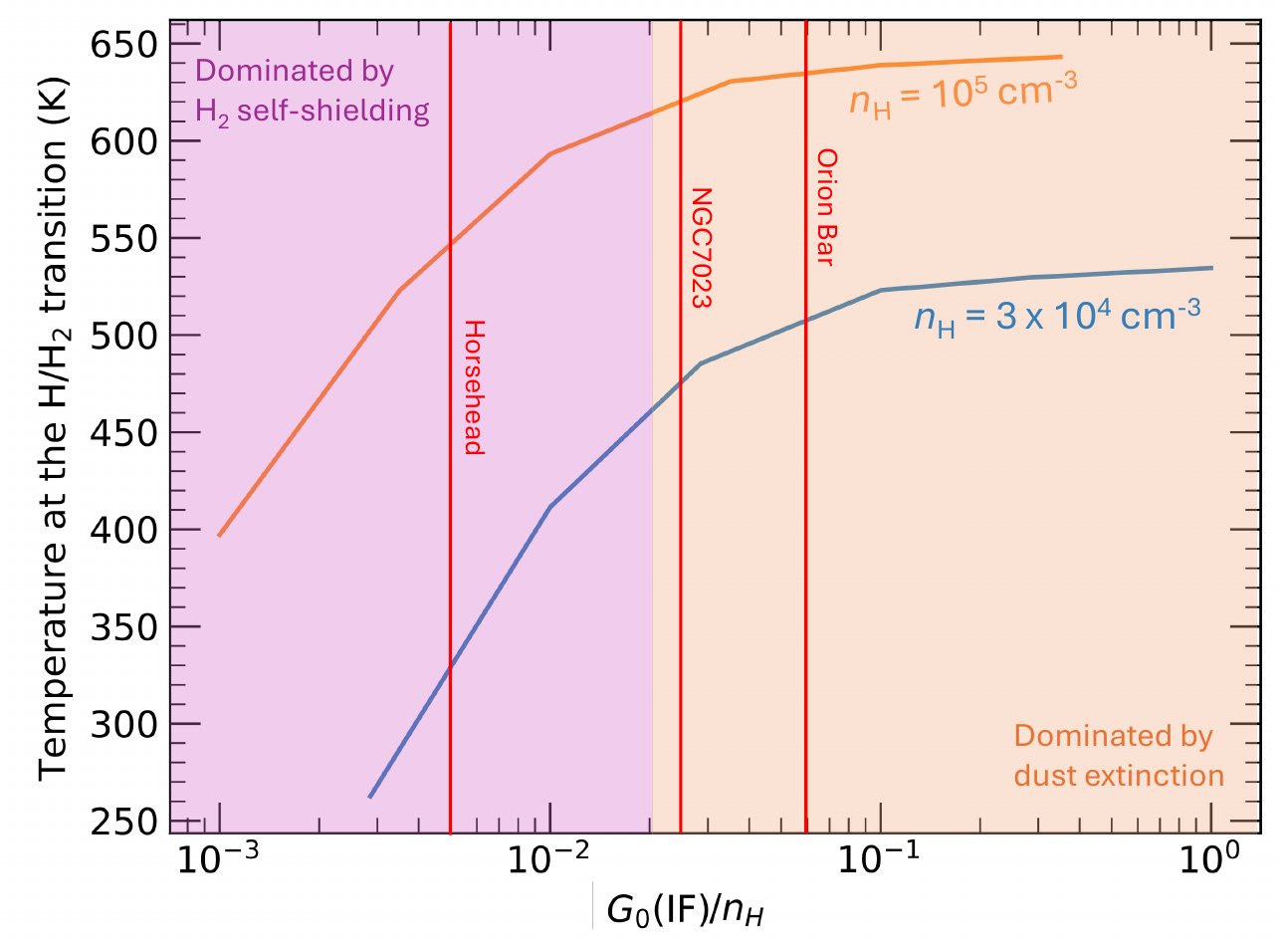}
			\caption{Gas temperature predicted at the H/H$_2$ transition ($x$(H)$=x$(\molh)) as a function of the ratio $G_0$(IF)/$n_{\rm H}$ for PDR models with densities of $n_{\rm H} = 3\times 10^4$~cm$^{-3}$ and $n_{\rm H} = 10^5$~cm$^{-3}$.}
			\label{fig:temp_h_h2_GOn}
		\end{figure}
		Figure \ref{fig:temp_h_h2_GOn} shows the variation of the temperature at the H/H$_2$ transition as a function of the ratio $G_0(\text{IF})/n_{\rm H}$ for models at a density $n_{\rm H} = 3 \times 10^4$~cm$^{-3}$ and $n_{\rm H} = 10^5$~cm$^{-3}$. We find that when the UV extinction shifts from \molh self-shielding to dust extinction, the temperature no longer depends on $G_0$(IF), consistent with the similar temperatures we find for NGC7023 and the Orion Bar. 
		
		\begin{figure}
			\centering
			\includegraphics[width=\linewidth]{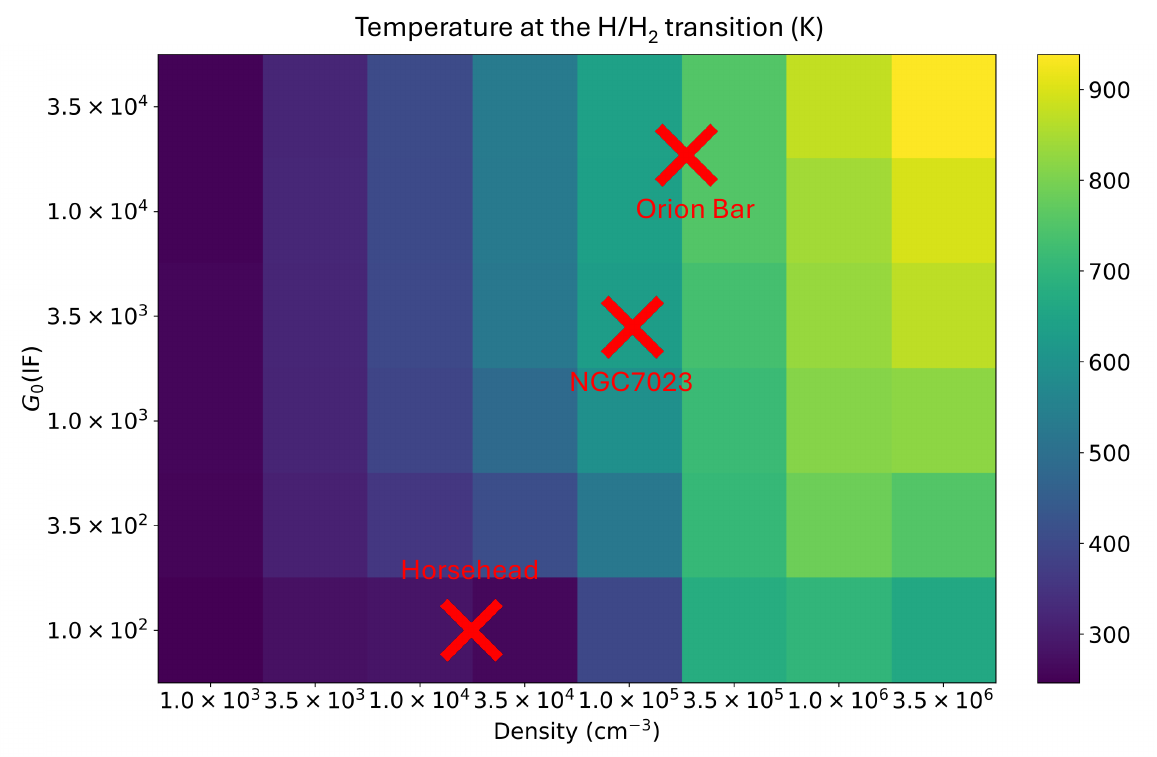}
			\caption{Gas temperature predicted at the H/H$_2$ transition ($x$(H)$=x$(\molh)) as a function of the gas density $n_{\rm H}$ and the intensity of the incident UV field $G_0$(IF).}
			\label{fig:temp_h_h2}
		\end{figure}
		Figure \ref{fig:temp_h_h2} displays the variation of the temperature at the H/H$_2$ transition as a function of the gas density $n_{\rm H}$ and the intensity of the incident UV field $G_0$(IF). Again, we observe that the temperatures in NGC7023 and the Orion Bar are predicted to be very similar (around $T_{\rm gas} \sim 600-700$~K, consistent with this study), whereas stationary models cannot reproduce the high temperature observed in the Horsehead Nebula (the predicted temperature is around $T_{\rm gas} \sim 300$~K). 
		
		\begin{figure}
			\centering
			\includegraphics[width=\linewidth]{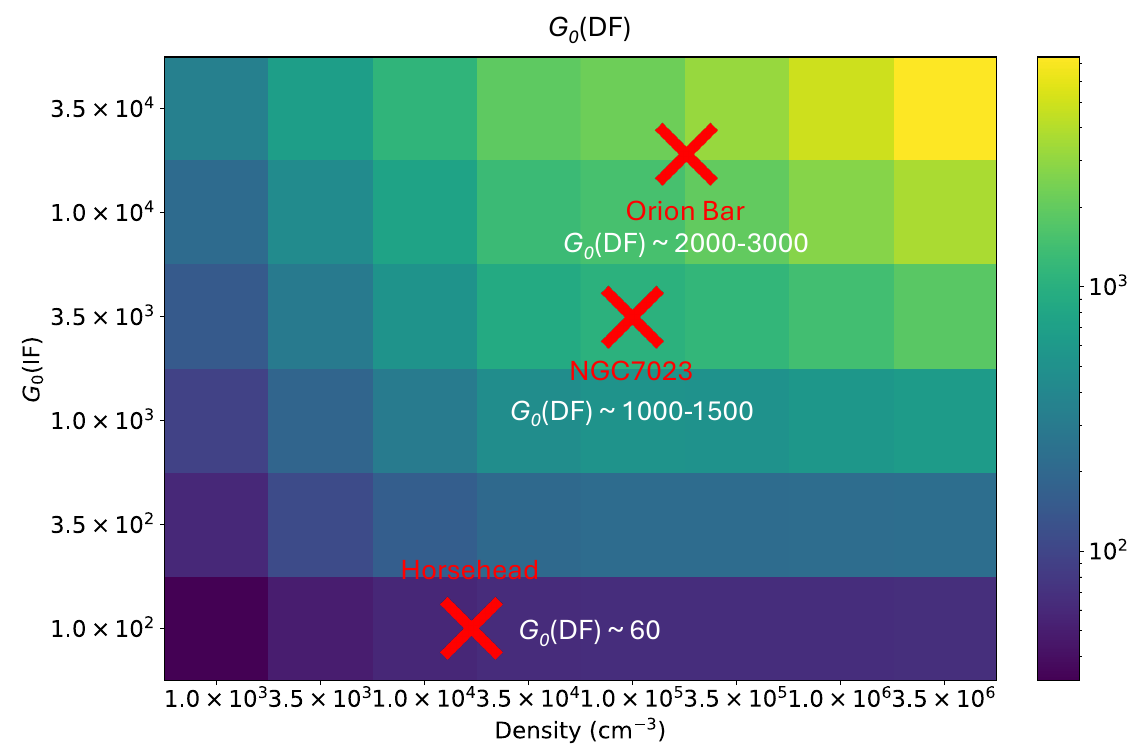}
			\caption{UV field intensity predicted at the H/H$_2$ transition ($x$(H)$=x$(\molh) $G_0$(DF) as a function of the gas density $n_{\rm H}$ and the intensity of the incident UV field $G_0$(IF).}
			\label{fig:GDF_h_h2}
		\end{figure}
		Figure \ref{fig:GDF_h_h2} displays the variation of the UV field intensity at the H/H$_2$ transition $G_0$(DF) as a function of the gas density $n_{\rm H}$ and the intensity of the incident UV field $G_0$(IF). This figure shows that for PDRs with $G_0(\text{IF})/n_{\rm H} < 0.02$, where the extinction of the UV field is dominated by H$_2$ self-shielding, the H/H$_2$ transition is very close to the edge of the PDR and thus the UV field intensity at the H/H$_2$ transition does not vary with gas density ($G_0$(DF) $\sim G_0$(IF)). On the contrary, for PDRs with $G_0(\text{IF})/n_{\rm H} > 0.02$, where the extinction of the UV field is dominated by dust extinction, the H/H$_2$ transition is located deeper into the PDR, and the higher the gas density, the higher the UV field intensity at the H/H$_2$ transition. However, the variation of $G_0$(DF) is much smaller than that of $G_0$(IF) (for a factor of 10 in $G_0$(IF), there is only a factor of 2 difference in $G_0$(DF)). This means that for PDRs with $G_0(\text{IF})/n_{\rm H} > 0.02$, the physical conditions at the H/H$_2$ transition are settled by the gas density and not the $G_0$(IF).

\begin{figure}
			\centering
			\includegraphics[width=\linewidth]{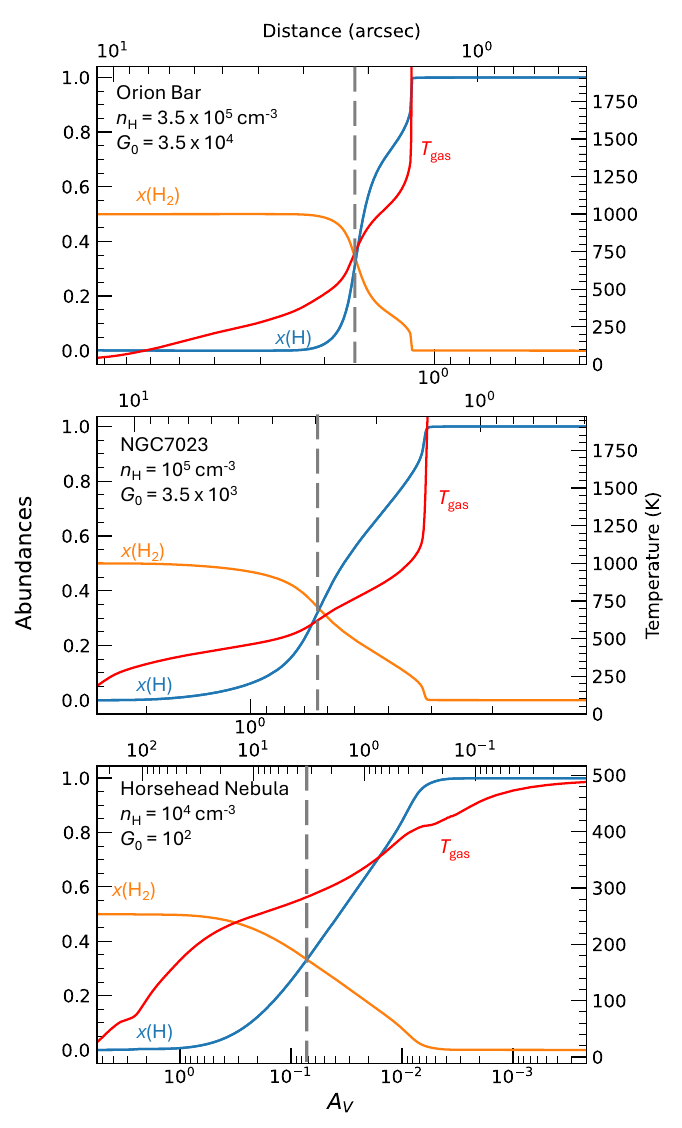}
			\caption{Abundances of H and \molh relative to H nuclei and gas temperature as a function of visual extinction ($A_V$) and distance to the edge of the PDR (in arcsec). The gray dashed line marks the position of the H/\molh transition.}
			\label{fig:temp_xh_xh2}
		\end{figure}

		Figure \ref{fig:temp_xh_xh2} displays the abundances of H and H$_2$ relative to H nuclei and the temperature as a function of the depth inside the PDR (in distance (arcsec) and visual extinction $A_V$) for isochoric models with representative parameters for the three PDRs. We observe that in the Horsehead Nebula, variations in temperature and abundance operate over larger spatial scales due to the lower gas density. 
		
		\onecolumn
		
		\section{Schematic view of the three PDRs}
\begin{figure*}[!h]
			\centering
			\includegraphics[width=0.35\linewidth]{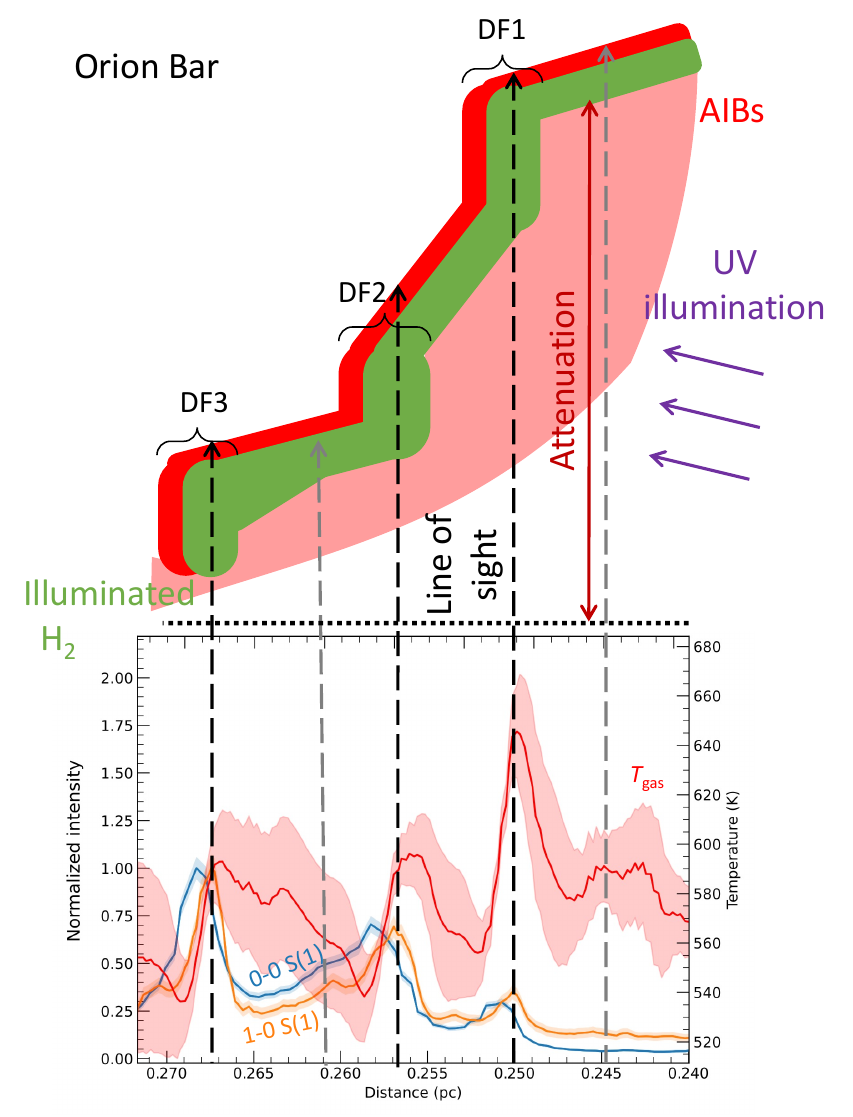}    \includegraphics[width=0.29\linewidth]{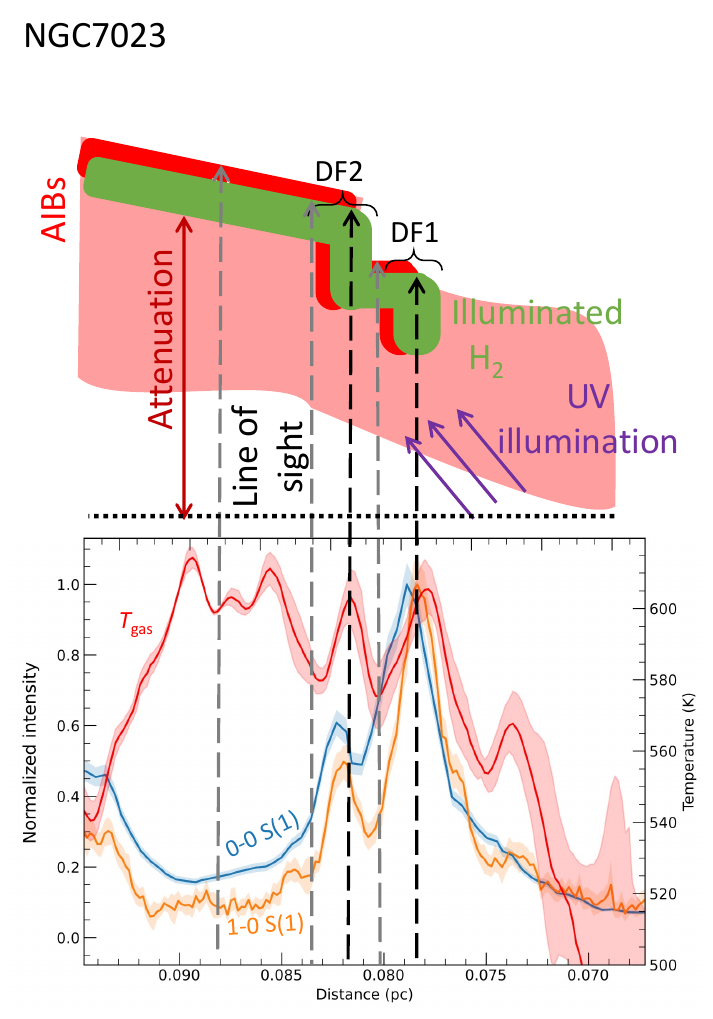}    \includegraphics[width=0.35\linewidth]{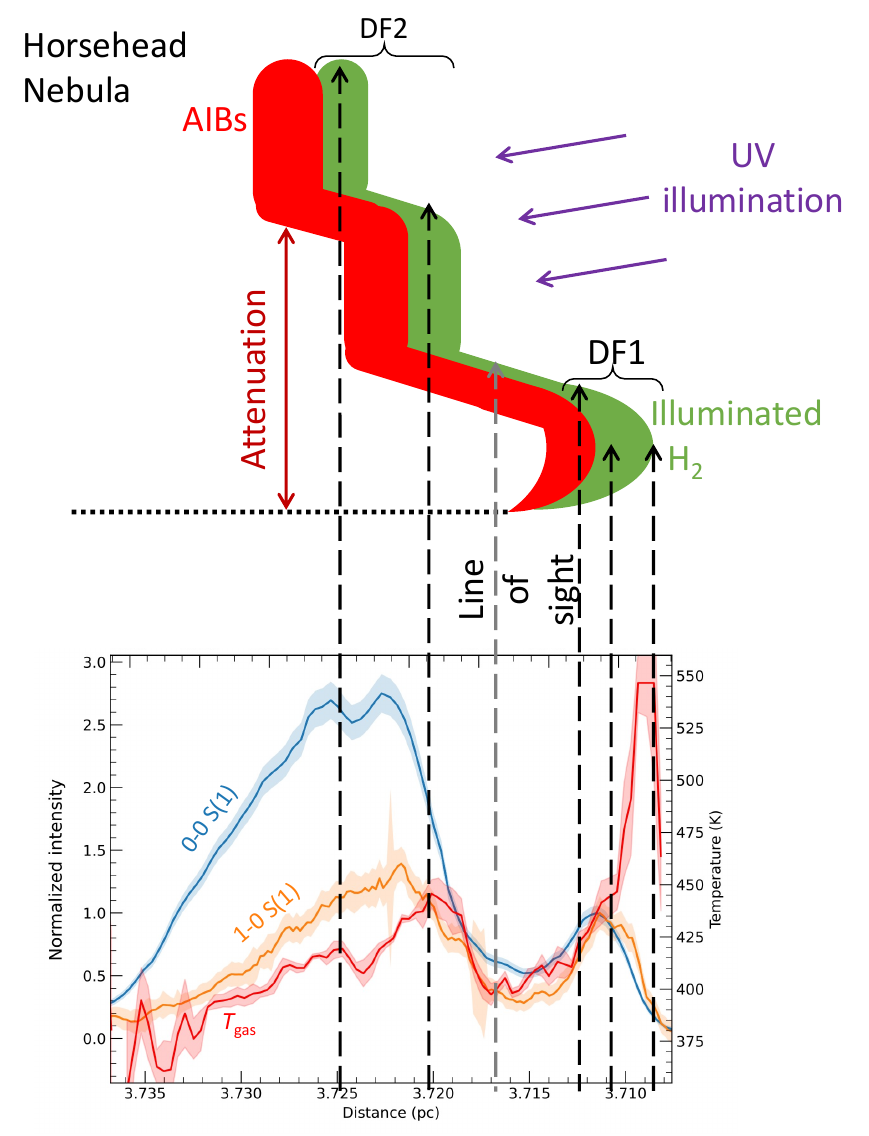}
			\caption{Comparison between the schematic view of the geometry of the Orion Bar (left panel),  NGC7023 (middle panel), and the Horsehead Nebula (right panel) together with the observed profile for \molh and temperature. The dark red corresponds to the peak of AIBs emission at the DF position, while the light red corresponds to the more extended AIBs emission in front of the DFs.}
			\label{fig:geometry}
		\end{figure*}

		Figure \ref{fig:geometry} shows a schematic view of the three PDRs, which explains the rather constant temperature observed across the entire FOV. Indeed, the illuminated layer of \molh dominates the emission, and therefore \molh emission can only probe the temperature in this thin layer.
		
		\section{Intensities of \molh lines used in the fitting procedure for the Orion Bar, NGC7023, and the Horsehead Nebula.}
		Table \ref{tab:intensity_h2} presents the intensities of H$_2$ used in the fitting procedure. 
		
		\FloatBarrier
		\begin{table*}[h!]
			\centering
			\begin{tabular}{c|c|c|c|c|c|c}
				Wavelength ($\mu$m)& Line &  $I$ ($\times 10^{-4}$) &  $I_{\rm corr} $  ($\times 10^{-4}$)  &  $I$  ($\times 10^{-4}$) &  $I_{\rm corr}$  ($\times 10^{-4}$) &  $I$  ($\times 10^{-6}$)    \\
				\hline\hline
				&  &  \multicolumn{2}{c|}{Orion Bar} & \multicolumn{2}{|c|}{NGC7023} & Horsehead Nebula \\
				\hline\hline
				
				17.037 & $0-0$ S(1)& $14.9 \pm 	0.2$ & 	$18.8 \pm 0.3$   & $5.79 \pm	0.04 $	& $6.32 \pm 0.05$ & $ 33.3\pm 0.2$   \\
				12.280 & $0-0$ S(2)&$12.0 \pm 	0.3$ &	$15.0 \pm 0.4$    & $ 5.77\pm	0.03$ &	$6.27 \pm 0.03$  &$40.3 \pm0.3$ \\
				9.666 &$0-0$ S(3)  & $21.4 \pm 	0.4 $&	$33.7 \pm 0.7$    & $14.4 \pm 0.1$ & 	$17.1\pm 0.2$ &$71.6\pm 0.4$  \\
				8.026 &$0-0$ S(4)  &$11.4 \pm 	0.4 $&	$14.2 \pm 	0.5$  & $5.69 \pm	0.04$ 	&$6.20 \pm	0.05$ &$20.8\pm 0.5$ \\
				6.910 & $0-0$ S(5) &$17.9 \pm	0.3$ &	$22.1 \pm	0.3$  &  $10.2 \pm	0.09$ &	$11.0\pm	0.09$ & $29.9\pm 0.6$\\
				6.109 & $0-0$ S(6) &$2.77 \pm	0.06$ &	$3.47 \pm 0.08$   &   $ 1.62 \pm	0.05$ &	$1.77\pm	0.06$&$ 6.2 \pm1.0 $   \\
				5.512 & $0-0$ S(7) &$4.9 \pm	0.3$ &	$6.3\pm		0.4$      &   $2.87 \pm	0.03$ 	& $3.14\pm	0.03$&$ 8.9\pm 1.2 $  \\
				5.053 & $0-0$ S(8) &$1.04 	\pm 0.02$ &	$1.35\pm	0.2$  &   $0.55 \pm	0.02$ &	$0.61\pm	0.02$ &$2.5\pm 0.2$ \\
				4.695 & $0-0$ S(9) & $2.47 \pm 	0.05$ 	&$3.26 \pm	0.07$ &  $1.16 \pm	0.04$ &	$1.29 \pm0.05$ &$4.3\pm 0.3$\\
				4.410 & $0-0$ S(10)&$0.58 \pm 0.03$ &	$7.77\pm	0.03$ & $0.216 \pm	0.007$ &	$0.242\pm 0.008$ &$      1.1\pm 0.3 $\\
				4.181 & $0-0$ S(11)&$1.45 \pm	0.06$ &	$1.98\pm	0.08$ &  $0.646\pm	0.008$ &	$0.728\pm	0.009$  &  $1.90\pm 0.07 $  \\
				3.996 & $0-0$ S(12)&$0.37 \pm	0.02$ &	$0.52\pm 0.02$    & $0.19 \pm	0.01$&	$0.21\pm	0.02$ & $0.59 \pm0.07$	\\
				3.846 &$0-0$ S(13) &$0.93 \pm 	0.04$ &	$1.32\pm	0.06$ &    $0.36\pm 0.06$& $0.42\pm 0.06$ &$0.89 \pm0.14$ \\ \hline
				
			\end{tabular}
			\caption{Intensities (in erg cm$^{-2}$ s$^{-1}$ sr$^{-1}$) of \molh lines used in the fitting procedure in the Orion Bar, NGC7023, and the Horsehead Nebula, not corrected ($I$) and corrected for extinction ($I_{\rm corr}$). The uncertainties are only the fitting error. Calibration effects are not accounted for, and the associated uncertainties can reach 5\% (as used in the excitation diagrams).}
			\label{tab:intensity_h2}
		\end{table*}
		\section{Data fusion}
		\label{appendix:fusion}
		Each astronomical instrument has inherent limitations resulting from its design, leading to trade-offs between spatial and spectral performance. This is particularly the case for JWST MIRI, where the Imager provides high spatial resolution but limited spectral information, while the MRS spectrometer offers high spectral resolution at a lower spatial resolution. In addition, in the infrared regime, the PSF broadens significantly at longer wavelengths. As a consequence, it is common practice to degrade all observations to the lowest spatial resolution in order to perform consistent scientific analyses.
		
		The data fusion method, based on a detailed instrumental modeling of both the MRS and the Imager, enables the reconstruction of a measurement that combines the strengths of each instrument. In this framework, we reconstruct a deconvolved hyperspectral data cube that achieves the spectral precision of the MRS together with the spatial resolution of the Imager ($\sim 0.3$"). The full method of data fusion is explained in Monnier et al. in prep. Furthermore, the fusion method enhances the spatial contrast between emission lines and the underlying continuum, enabling more accurate analysis of their morphological variations.
		
\begin{figure*}[!h]
			\centering
			\includegraphics[width=\linewidth]{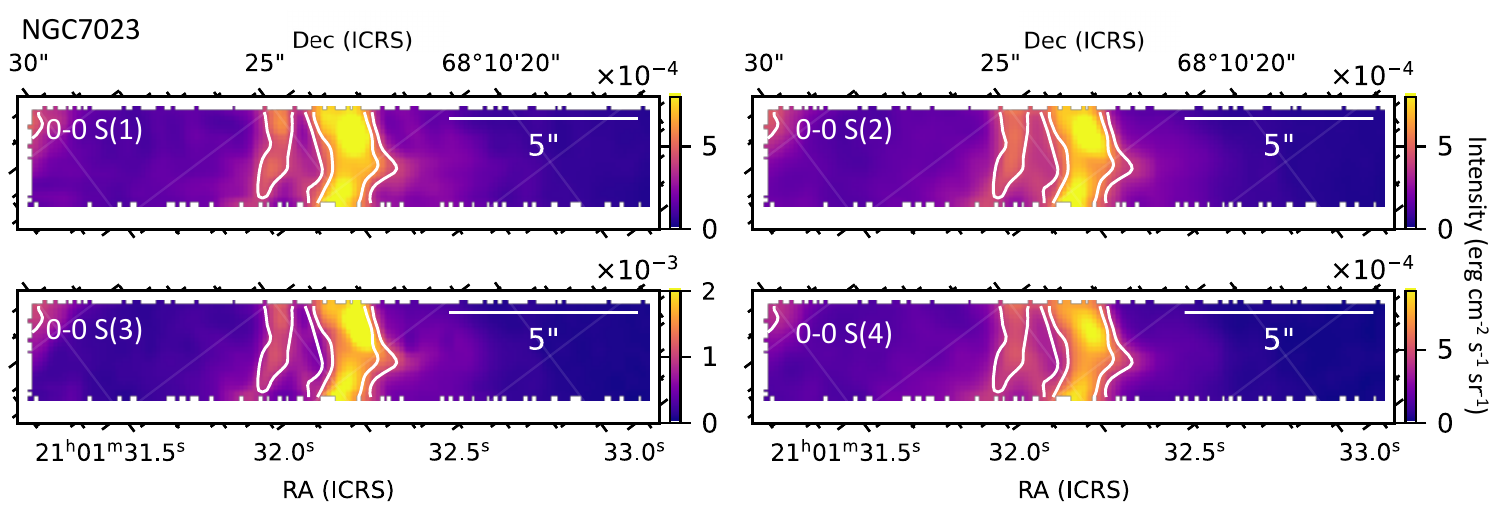}
			\caption{Maps of some \molh rotational line emission and the rovibrational 1–0 S(1) and 2–1 S(1) lines emission obtained with MIRI/MRS and NIRSpec using data fusion. White contours are from the 0–0 S(1) line emission. The line intensities are not corrected for extinction. The illuminating star is located on the right in all panels.}
			\label{fig:H2mapsfusion}
		\end{figure*}

\begin{figure}[!h]
			\centering
			\includegraphics[width=\linewidth]{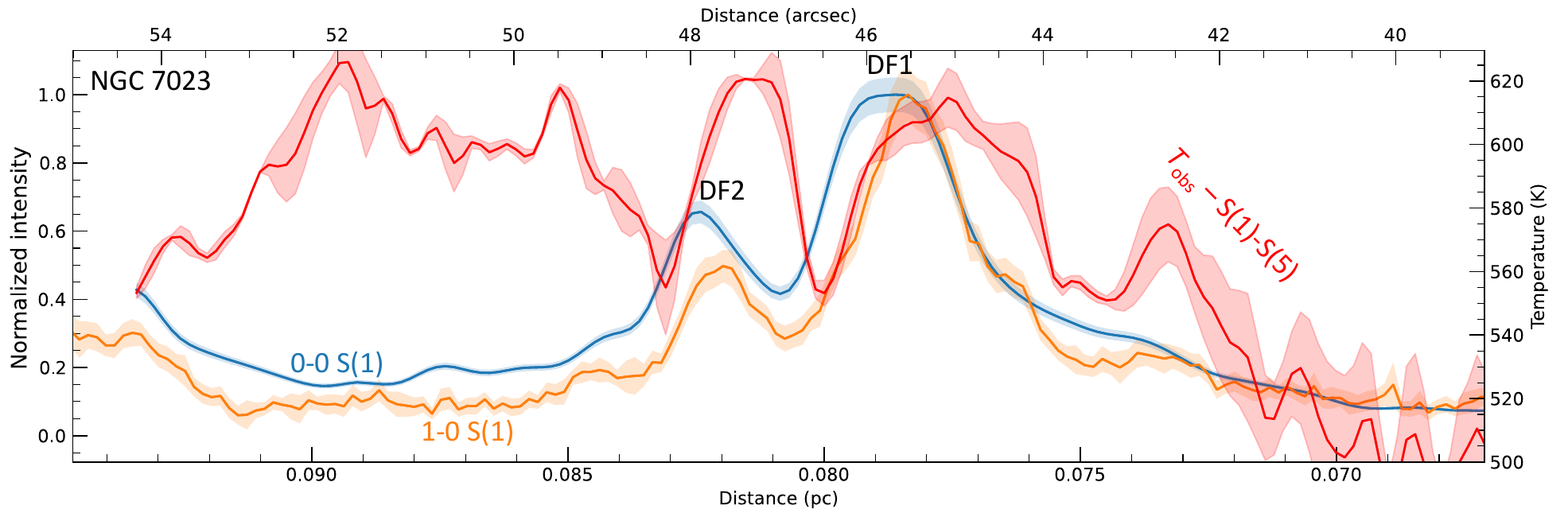}
			\caption{Gas temperature profile derived from a single-component fit of the first five observed \molh lines (S(1)-S(5)) using data from the fusion method, without the S(3) line, compared to the \molh 0–0 S(1) and 1–0 S(1) line emission profile in NGC7023 as a function of the distance to the illuminating star (pc and arcsec). The decrease in temperature in the different \molh filaments is better resolved than in the non-fused data.}
			\label{fig:cutfusion}
		\end{figure}

		Figure \ref{fig:H2mapsfusion} shows the maps of the first four observed rotational lines of \molh using this method. This figure shows how we can better resolve the filaments, especially for the $0-0$ S(1) line. At this wavelength $\lambda(0-0$ S(1))$=17.03$~$\mu$m, before the fusion, we can only produce maps with a spatial resolution
		of 0.7". Hence, the spatial resolution is increased by more than a factor of two for this line. Figure \ref{fig:cutfusion} presents the observed temperature profile along the cuts presented in Fig. \ref{fig:FOVbis} using the data from the fusion method. The decrease in temperature in each DF is also better resolved when using this method. 
		
	\end{appendix}

\end{document}